\documentclass[aps,reprint,twocolumn,10pt,superscriptaddress,english,showpacs,longbibliography,nofootinbib]{revtex4-1}
\usepackage{amsmath}
\usepackage{amssymb}
\usepackage{amsthm}
\usepackage{amsfonts}
\usepackage{listings}
\usepackage{longtable}
\usepackage{enumerate}
\usepackage{latexsym}
\usepackage{color}

\usepackage{multirow}
\usepackage[figuresright]{rotating}

\usepackage[table]{xcolor}
\usepackage{setspace} 
\usepackage{blindtext}

\usepackage{array,etoolbox}
\usepackage[titles]{tocloft}
\preto\tabular{\setcounter{magicrownumbers}{0}}
\newcounter{magicrownumbers}

\newcommand{\icsdwebshort}[1]{\href{https://www.topologicalquantumchemistry.fr/\#/detail/#1}{#1}}
\newcommand{\icsdweb}[1]{\href{https://www.topologicalquantumchemistry.fr/\#/detail/#1}{ICSD #1}}

\newcommand{\webTQC}{\href{https://www.topologicalquantumchemistry.fr/}{Topological Quantum Chemistry website}}
\newcommand{\webMTQC}{\href{https://www.topologicalquantumchemistry.fr/magnetic}{Topological Magnetic Materials website}}

\newcommand{\bcsidweblong}[1]{\href{https://www.topologicalquantumchemistry.fr/magnetic/index.html?BCSID=#1}{BCSID #1}}

\definecolor{orange}{rgb}{1,0.5,0}
\definecolor{Red}{RGB}{255,204,204}
\definecolor{Blue}{RGB}{153,204,255}
\definecolor{Green}{RGB}{153,255,153}
\definecolor{Yellow}{RGB}{255,255,153}
\definecolor{Black}{RGB}{224,224,224}

\usepackage{bm}
\usepackage{hyperref}
\hypersetup{
 pdfnewwindow=true, colorlinks=true,
 linkcolor=blue, anchorcolor=blue,
 citecolor=blue, filecolor=blue,
 menucolor=blue, urlcolor=blue}
\definecolor{red}{rgb}{1,0,0}
\definecolor{blue}{rgb}{0,0,1}
\definecolor{black}{rgb}{0,0,0}

\usepackage{psfrag}

\usepackage{bm}
\usepackage{graphicx}

\RequirePackage[normalem]{ulem} 
\RequirePackage{color}\definecolor{RED}{rgb}{1,0,0}\definecolor{BLUE}{rgb}{0,0,1} 

\def\ie{{\it i.e.},\ }

\input{epsf}

\begin{document}


\newcommand{\sgsymb}[1]{\ifnum#1=1
$P1$\else
\ifnum#1=2
$P\bar{1}$\else
\ifnum#1=3
$P2$\else
\ifnum#1=4
$P2_1$\else
\ifnum#1=5
$C2$\else
\ifnum#1=6
$Pm$\else
\ifnum#1=7
$Pc$\else
\ifnum#1=8
$Cm$\else
\ifnum#1=9
$Cc$\else
\ifnum#1=10
$P2/m$\else
\ifnum#1=11
$P2_1/m$\else
\ifnum#1=12
$C2/m$\else
\ifnum#1=13
$P2/c$\else
\ifnum#1=14
$P2_1/c$\else
\ifnum#1=15
$C2/c$\else
\ifnum#1=16
$P222$\else
\ifnum#1=17
$P222_1$\else
\ifnum#1=18
$P2_12_12$\else
\ifnum#1=19
$P2_12_12_1$\else
\ifnum#1=20
$C222_1$\else
\ifnum#1=21
$C222$\else
\ifnum#1=22
$F222$\else
\ifnum#1=23
$I222$\else
\ifnum#1=24
$I2_12_12_1$\else
\ifnum#1=25
$Pmm2$\else
\ifnum#1=26
$Pmc2_1$\else
\ifnum#1=27
$Pcc2$\else
\ifnum#1=28
$Pma2$\else
\ifnum#1=29
$Pca2_1$\else
\ifnum#1=30
$Pnc2$\else
\ifnum#1=31
$Pmn2_1$\else
\ifnum#1=32
$Pba2$\else
\ifnum#1=33
$Pna2_1$\else
\ifnum#1=34
$Pnn2$\else
\ifnum#1=35
$Cmm2$\else
\ifnum#1=36
$Cmc2_1$\else
\ifnum#1=37
$Ccc2$\else
\ifnum#1=38
$Amm2$\else
\ifnum#1=39
$Aem2$\else
\ifnum#1=40
$Ama2$\else
\ifnum#1=41
$Aea2$\else
\ifnum#1=42
$Fmm2$\else
\ifnum#1=43
$Fdd2$\else
\ifnum#1=44
$Imm2$\else
\ifnum#1=45
$Iba2$\else
\ifnum#1=46
$Ima2$\else
\ifnum#1=47
$Pmmm$\else
\ifnum#1=48
$Pnnn$\else
\ifnum#1=49
$Pccm$\else
\ifnum#1=50
$Pban$\else
\ifnum#1=51
$Pmma$\else
\ifnum#1=52
$Pnna$\else
\ifnum#1=53
$Pmna$\else
\ifnum#1=54
$Pcca$\else
\ifnum#1=55
$Pbam$\else
\ifnum#1=56
$Pccn$\else
\ifnum#1=57
$Pbcm$\else
\ifnum#1=58
$Pnnm$\else
\ifnum#1=59
$Pmmn$\else
\ifnum#1=60
$Pbcn$\else
\ifnum#1=61
$Pbca$\else
\ifnum#1=62
$Pnma$\else
\ifnum#1=63
$Cmcm$\else
\ifnum#1=64
$Cmce$\else
\ifnum#1=65
$Cmmm$\else
\ifnum#1=66
$Cccm$\else
\ifnum#1=67
$Cmme$\else
\ifnum#1=68
$Ccce$\else
\ifnum#1=69
$Fmmm$\else
\ifnum#1=70
$Fddd$\else
\ifnum#1=71
$Immm$\else
\ifnum#1=72
$Ibam$\else
\ifnum#1=73
$Ibca$\else
\ifnum#1=74
$Imma$\else
\ifnum#1=75
$P4$\else
\ifnum#1=76
$P4_1$\else
\ifnum#1=77
$P4_2$\else
\ifnum#1=78
$P4_3$\else
\ifnum#1=79
$I4$\else
\ifnum#1=80
$I4_1$\else
\ifnum#1=81
$P\bar{4}$\else
\ifnum#1=82
$I\bar{4}$\else
\ifnum#1=83
$P4/m$\else
\ifnum#1=84
$P4_2/m$\else
\ifnum#1=85
$P4/n$\else
\ifnum#1=86
$P4_2/n$\else
\ifnum#1=87
$I4/m$\else
\ifnum#1=88
$I4_1/a$\else
\ifnum#1=89
$P422$\else
\ifnum#1=90
$P42_12$\else
\ifnum#1=91
$P4_122$\else
\ifnum#1=92
$P4_12_12$\else
\ifnum#1=93
$P4_222$\else
\ifnum#1=94
$P4_22_12$\else
\ifnum#1=95
$P4_322$\else
\ifnum#1=96
$P4_32_12$\else
\ifnum#1=97
$I422$\else
\ifnum#1=98
$I4_122$\else
\ifnum#1=99
$P4mm$\else
\ifnum#1=100
$P4bm$\else
\ifnum#1=101
$P4_2cm$\else
\ifnum#1=102
$P4_2nm$\else
\ifnum#1=103
$P4cc$\else
\ifnum#1=104
$P4nc$\else
\ifnum#1=105
$P4_2mc$\else
\ifnum#1=106
$P4_2bc$\else
\ifnum#1=107
$I4mm$\else
\ifnum#1=108
$I4cm$\else
\ifnum#1=109
$I4_1md$\else
\ifnum#1=110
$I4_1cd$\else
\ifnum#1=111
$P\bar{4}2m$\else
\ifnum#1=112
$P\bar{4}2c$\else
\ifnum#1=113
$P\bar{4}2_1m$\else
\ifnum#1=114
$P\bar{4}2_1c$\else
\ifnum#1=115
$P\bar{4}m2$\else
\ifnum#1=116
$P\bar{4}c2$\else
\ifnum#1=117
$P\bar{4}b2$\else
\ifnum#1=118
$P\bar{4}n2$\else
\ifnum#1=119
$I\bar{4}m2$\else
\ifnum#1=120
$I\bar{4}c2$\else
\ifnum#1=121
$I\bar{4}2m$\else
\ifnum#1=122
$I\bar{4}2d$\else
\ifnum#1=123
$P4/mmm$\else
\ifnum#1=124
$P4/mcc$\else
\ifnum#1=125
$P4/nbm$\else
\ifnum#1=126
$P4/nnc$\else
\ifnum#1=127
$P4/mbm$\else
\ifnum#1=128
$P4/mnc$\else
\ifnum#1=129
$P4/nmm$\else
\ifnum#1=130
$P4/ncc$\else
\ifnum#1=131
$P4_2/mmc$\else
\ifnum#1=132
$P4_2/mcm$\else
\ifnum#1=133
$P4_2/nbc$\else
\ifnum#1=134
$P4_2/nnm$\else
\ifnum#1=135
$P4_2/mbc$\else
\ifnum#1=136
$P4_2/mnm$\else
\ifnum#1=137
$P4_2/nmc$\else
\ifnum#1=138
$P4_2/ncm$\else
\ifnum#1=139
$I4/mmm$\else
\ifnum#1=140
$I4/mcm$\else
\ifnum#1=141
$I4_1/amd$\else
\ifnum#1=142
$I4_1/acd$\else
\ifnum#1=143
$P3$\else
\ifnum#1=144
$P3_1$\else
\ifnum#1=145
$P3_2$\else
\ifnum#1=146
$R3$\else
\ifnum#1=147
$P\bar{3}$\else
\ifnum#1=148
$R\bar{3}$\else
\ifnum#1=149
$P312$\else
\ifnum#1=150
$P321$\else
\ifnum#1=151
$P3_112$\else
\ifnum#1=152
$P3_121$\else
\ifnum#1=153
$P3_212$\else
\ifnum#1=154
$P3_221$\else
\ifnum#1=155
$R32$\else
\ifnum#1=156
$P3m1$\else
\ifnum#1=157
$P31m$\else
\ifnum#1=158
$P3c1$\else
\ifnum#1=159
$P31c$\else
\ifnum#1=160
$R3m$\else
\ifnum#1=161
$R3c$\else
\ifnum#1=162
$P\bar{3}1m$\else
\ifnum#1=163
$P\bar{3}1c$\else
\ifnum#1=164
$P\bar{3}m1$\else
\ifnum#1=165
$P\bar{3}c1$\else
\ifnum#1=166
$R\bar{3}m$\else
\ifnum#1=167
$R\bar{3}c$\else
\ifnum#1=168
$P6$\else
\ifnum#1=169
$P6_1$\else
\ifnum#1=170
$P6_5$\else
\ifnum#1=171
$P6_2$\else
\ifnum#1=172
$P6_4$\else
\ifnum#1=173
$P6_3$\else
\ifnum#1=174
$P\bar{6}$\else
\ifnum#1=175
$P6/m$\else
\ifnum#1=176
$P6_3/m$\else
\ifnum#1=177
$P622$\else
\ifnum#1=178
$P6_122$\else
\ifnum#1=179
$P6_522$\else
\ifnum#1=180
$P6_222$\else
\ifnum#1=181
$P6_422$\else
\ifnum#1=182
$P6_322$\else
\ifnum#1=183
$P6mm$\else
\ifnum#1=184
$P6cc$\else
\ifnum#1=185
$P6_3cm$\else
\ifnum#1=186
$P6_3mc$\else
\ifnum#1=187
$P\bar{6}m2$\else
\ifnum#1=188
$P\bar{6}c2$\else
\ifnum#1=189
$P\bar{6}2m$\else
\ifnum#1=190
$P\bar{6}2c$\else
\ifnum#1=191
$P6/mmm$\else
\ifnum#1=192
$P6/mcc$\else
\ifnum#1=193
$P6_3/mcm$\else
\ifnum#1=194
$P6_3/mmc$\else
\ifnum#1=195
$P23$\else
\ifnum#1=196
$F23$\else
\ifnum#1=197
$I23$\else
\ifnum#1=198
$P2_13$\else
\ifnum#1=199
$I2_13$\else
\ifnum#1=200
$Pm\bar{3}$\else
\ifnum#1=201
$Pn\bar{3}$\else
\ifnum#1=202
$Fm\bar{3}$\else
\ifnum#1=203
$Fd\bar{3}$\else
\ifnum#1=204
$Im\bar{3}$\else
\ifnum#1=205
$Pa\bar{3}$\else
\ifnum#1=206
$Ia\bar{3}$\else
\ifnum#1=207
$P432$\else
\ifnum#1=208
$P4_232$\else
\ifnum#1=209
$F432$\else
\ifnum#1=210
$F4_132$\else
\ifnum#1=211
$I432$\else
\ifnum#1=212
$P4_332$\else
\ifnum#1=213
$P4_132$\else
\ifnum#1=214
$I4_132$\else
\ifnum#1=215
$P\bar{4}3m$\else
\ifnum#1=216
$F\bar{4}3m$\else
\ifnum#1=217
$I\bar{4}3m$\else
\ifnum#1=218
$P\bar{4}3n$\else
\ifnum#1=219
$F\bar{4}3c$\else
\ifnum#1=220
$I\bar{4}3d$\else
\ifnum#1=221
$Pm\bar{3}m$\else
\ifnum#1=222
$Pn\bar{3}n$\else
\ifnum#1=223
$Pm\bar{3}n$\else
\ifnum#1=224
$Pn\bar{3}m$\else
\ifnum#1=225
$Fm\bar{3}m$\else
\ifnum#1=226
$Fm\bar{3}c$\else
\ifnum#1=227
$Fd\bar{3}m$\else
\ifnum#1=228
$Fd\bar{3}c$\else
\ifnum#1=229
$Im\bar{3}m$\else
\ifnum#1=230
$Ia\bar{3}d$\else
{\color{red}{Invalid SG number}}
\fi
\fi
\fi
\fi
\fi
\fi
\fi
\fi
\fi
\fi
\fi
\fi
\fi
\fi
\fi
\fi
\fi
\fi
\fi
\fi
\fi
\fi
\fi
\fi
\fi
\fi
\fi
\fi
\fi
\fi
\fi
\fi
\fi
\fi
\fi
\fi
\fi
\fi
\fi
\fi
\fi
\fi
\fi
\fi
\fi
\fi
\fi
\fi
\fi
\fi
\fi
\fi
\fi
\fi
\fi
\fi
\fi
\fi
\fi
\fi
\fi
\fi
\fi
\fi
\fi
\fi
\fi
\fi
\fi
\fi
\fi
\fi
\fi
\fi
\fi
\fi
\fi
\fi
\fi
\fi
\fi
\fi
\fi
\fi
\fi
\fi
\fi
\fi
\fi
\fi
\fi
\fi
\fi
\fi
\fi
\fi
\fi
\fi
\fi
\fi
\fi
\fi
\fi
\fi
\fi
\fi
\fi
\fi
\fi
\fi
\fi
\fi
\fi
\fi
\fi
\fi
\fi
\fi
\fi
\fi
\fi
\fi
\fi
\fi
\fi
\fi
\fi
\fi
\fi
\fi
\fi
\fi
\fi
\fi
\fi
\fi
\fi
\fi
\fi
\fi
\fi
\fi
\fi
\fi
\fi
\fi
\fi
\fi
\fi
\fi
\fi
\fi
\fi
\fi
\fi
\fi
\fi
\fi
\fi
\fi
\fi
\fi
\fi
\fi
\fi
\fi
\fi
\fi
\fi
\fi
\fi
\fi
\fi
\fi
\fi
\fi
\fi
\fi
\fi
\fi
\fi
\fi
\fi
\fi
\fi
\fi
\fi
\fi
\fi
\fi
\fi
\fi
\fi
\fi
\fi
\fi
\fi
\fi
\fi
\fi
\fi
\fi
\fi
\fi
\fi
\fi
\fi
\fi
\fi
\fi
\fi
\fi
\fi
\fi
\fi
\fi
\fi
\fi
\fi
\fi
\fi
\fi
\fi
\fi
\fi
\fi
\fi
\fi
\fi
\fi}

\newcommand{\sgsymbnum}[1]{SG #1 (\sgsymb{#1})}

\newcommand{\TQCDBNbrICSDs}{73,234}
\newcommand{\TQCDBNbrUniqueMaterials}{38,298}

\newcommand{\TQCDBNbrICSDsTrivial}{34,013}
\newcommand{\TQCDBNbrICSDsTrivialPercent}{46.44\%}

\newcommand{\TQCDBNbrMaterialsTrivial}{18,133}
\newcommand{\TQCDBNbrMaterialsTrivialPercent}{47.35\%}

\newcommand{\TQCDBNbrICSDsFeOAI}{957}
\newcommand{\TQCDBNbrICSDsFeOAIPercent}{2.81\%}

\newcommand{\TQCDBNbrMaterialsFeOAI}{638}
\newcommand{\TQCDBNbrMaterialsFeOAIPercent}{3.52\%}

\newcommand{\TQCDBNbrICSDsFeOAIIndirectGap}{738}
\newcommand{\TQCDBNbrICSDsFeOAIIndirectGapPercent}{100.00\%}

\newcommand{\TQCDBNbrMaterialsFeOAIIndirectGap}{475}
\newcommand{\TQCDBNbrMaterialsFeOAIIndirectGapPercent}{2.62\%}


\newcommand{\TQCDBNbrFEOAIPerSG}[1]{\ifnum#1=2
\begin{tabular}{c}108 \\ (92)\end{tabular}\else
\ifnum#1=5
\begin{tabular}{c}6 \\ (6)\end{tabular} \else
\ifnum#1=12
\begin{tabular}{c}108 \\ (81)\end{tabular} \else
\ifnum#1=13
\begin{tabular}{c}1 \\ (1)\end{tabular} \else
\ifnum#1=14
\begin{tabular}{c}252 \\ (167)\end{tabular} \else
\ifnum#1=15
\begin{tabular}{c}44 \\ (34)\end{tabular} \else
\ifnum#1=20
\begin{tabular}{c}1 \\ (1)\end{tabular} \else
\ifnum#1=38
\begin{tabular}{c}2 \\ (2)\end{tabular} \else
\ifnum#1=41
\begin{tabular}{c}2 \\ (2)\end{tabular} \else
\ifnum#1=43
\begin{tabular}{c}1 \\ (1)\end{tabular} \else
\ifnum#1=45
\begin{tabular}{c}1 \\ (1)\end{tabular} \else
\ifnum#1=55
\begin{tabular}{c}43 \\ (35)\end{tabular} \else
\ifnum#1=58
\begin{tabular}{c}17 \\ (9)\end{tabular} \else
\ifnum#1=60
\begin{tabular}{c}4 \\ (3)\end{tabular} \else
\ifnum#1=61
\begin{tabular}{c}45 \\ (25)\end{tabular} \else
\ifnum#1=63
\begin{tabular}{c}1 \\ (1)\end{tabular} \else
\ifnum#1=64
\begin{tabular}{c}50 \\ (21)\end{tabular} \else
\ifnum#1=65
\begin{tabular}{c}1 \\ (1)\end{tabular} \else
\ifnum#1=69
\begin{tabular}{c}6 \\ (5)\end{tabular} \else
\ifnum#1=70
\begin{tabular}{c}12 \\ (11)\end{tabular} \else
\ifnum#1=71
\begin{tabular}{c}27 \\ (8)\end{tabular} \else
\ifnum#1=81
\begin{tabular}{c}1 \\ (1)\end{tabular} \else
\ifnum#1=82
\begin{tabular}{c}8 \\ (7)\end{tabular} \else
\ifnum#1=86
\begin{tabular}{c}4 \\ (2)\end{tabular} \else
\ifnum#1=87
\begin{tabular}{c}4 \\ (3)\end{tabular} \else
\ifnum#1=88
\begin{tabular}{c}7 \\ (3)\end{tabular} \else
\ifnum#1=91
\begin{tabular}{c}1 \\ (1)\end{tabular} \else
\ifnum#1=92
\begin{tabular}{c}3 \\ (1)\end{tabular} \else
\ifnum#1=102
\begin{tabular}{c}2 \\ (2)\end{tabular} \else
\ifnum#1=114
\begin{tabular}{c}1 \\ (1)\end{tabular} \else
\ifnum#1=118
\begin{tabular}{c}3 \\ (3)\end{tabular} \else
\ifnum#1=122
\begin{tabular}{c}1 \\ (1)\end{tabular} \else
\ifnum#1=126
\begin{tabular}{c}4 \\ (4)\end{tabular} \else
\ifnum#1=136
\begin{tabular}{c}17 \\ (9)\end{tabular} \else
\ifnum#1=137
\begin{tabular}{c}1 \\ (1)\end{tabular} \else
\ifnum#1=138
\begin{tabular}{c}1 \\ (1)\end{tabular} \else
\ifnum#1=139
\begin{tabular}{c}25 \\ (12)\end{tabular} \else
\ifnum#1=141
\begin{tabular}{c}4 \\ (2)\end{tabular} \else
\ifnum#1=142
\begin{tabular}{c}9 \\ (2)\end{tabular} \else
\ifnum#1=147
\begin{tabular}{c}1 \\ (1)\end{tabular} \else
\ifnum#1=148
\begin{tabular}{c}24 \\ (15)\end{tabular} \else
\ifnum#1=152
\begin{tabular}{c}1 \\ (1)\end{tabular} \else
\ifnum#1=162
\begin{tabular}{c}3 \\ (3)\end{tabular} \else
\ifnum#1=164
\begin{tabular}{c}2 \\ (2)\end{tabular} \else
\ifnum#1=165
\begin{tabular}{c}1 \\ (1)\end{tabular} \else
\ifnum#1=166
\begin{tabular}{c}18 \\ (10)\end{tabular} \else
\ifnum#1=167
\begin{tabular}{c}1 \\ (1)\end{tabular} \else
\ifnum#1=194
\begin{tabular}{c}29 \\ (6)\end{tabular} \else
\ifnum#1=197
\begin{tabular}{c}1 \\ (1)\end{tabular} \else
\ifnum#1=199
\begin{tabular}{c}4 \\ (3)\end{tabular} \else
\ifnum#1=202
\begin{tabular}{c}9 \\ (5)\end{tabular} \else
\ifnum#1=204
\begin{tabular}{c}5 \\ (1)\end{tabular} \else
\ifnum#1=205
\begin{tabular}{c}6 \\ (3)\end{tabular} \else
\ifnum#1=214
\begin{tabular}{c}3 \\ (3)\end{tabular} \else
\ifnum#1=215
\begin{tabular}{c}1 \\ (1)\end{tabular} \else
\ifnum#1=216
\begin{tabular}{c}4 \\ (3)\end{tabular} \else
\ifnum#1=217
\begin{tabular}{c}6 \\ (6)\end{tabular} \else
\ifnum#1=220
\begin{tabular}{c}1 \\ (1)\end{tabular} \else
\ifnum#1=225
\begin{tabular}{c}2 \\ (2)\end{tabular} \else
\ifnum#1=227
\begin{tabular}{c}6 \\ (5)\end{tabular} \else
\ifnum#1=229
\begin{tabular}{c}1 \\ (1)\end{tabular} \else
\ifnum#1=0
\begin{tabular}{c}957 \\ (638)\end{tabular} \else
{\color{red}{Invalid SG number}}
\fi
\fi
\fi
\fi
\fi
\fi
\fi
\fi
\fi
\fi
\fi
\fi
\fi
\fi
\fi
\fi
\fi
\fi
\fi
\fi
\fi
\fi
\fi
\fi
\fi
\fi
\fi
\fi
\fi
\fi
\fi
\fi
\fi
\fi
\fi
\fi
\fi
\fi
\fi
\fi
\fi
\fi
\fi
\fi
\fi
\fi
\fi
\fi
\fi
\fi
\fi
\fi
\fi
\fi
\fi
\fi
\fi
\fi
\fi
\fi
\fi
\fi
}

\title{Filling-Enforced Obstructed Atomic Insulators}

\author{Yuanfeng Xu}
\affiliation{Max Planck Institute of Microstructure Physics, 06120 Halle, Germany}

\author{Luis Elcoro}
\affiliation{Department of Condensed Matter Physics, University of the Basque Country UPV/EHU, Apartado 644, 48080 Bilbao, Spain}

\author{Zhi-Da Song}
\affiliation{Department of Physics, Princeton University, Princeton, New Jersey 08544, USA}

\author{M. G. Vergniory}
\affiliation{Donostia International Physics Center, P. Manuel de Lardizabal 4, 20018 Donostia-San Sebastian, Spain}
\affiliation{IKERBASQUE, Basque Foundation for Science, Bilbao, Spain}
\affiliation{Max Planck Institute for Chemical Physics of Solids, 01309 Dresden, Germany}

\author{Claudia Felser}
\affiliation{Max Planck Institute for Chemical Physics of Solids, 01309 Dresden, Germany}

\author{Stuart S. P. Parkin}
\affiliation{Max Planck Institute of Microstructure Physics, 06120 Halle, Germany}

\author{Nicolas Regnault}
\affiliation{Department of Physics, Princeton University, Princeton, New Jersey 08544, USA}
\affiliation{Laboratoire de Physique de l'Ecole normale sup\'{e}rieure, ENS, Universit\'{e} PSL, CNRS, Sorbonne Universit\'{e}, Universit\'{e} Paris-Diderot, Sorbonne Paris Cit\'{e}, 75005 Paris, France}

\author{Juan L. Ma\~nes}
\affiliation{Department of Condensed Matter Physics, University of the Basque Country UPV/EHU, Apartado 644, 48080 Bilbao, Spain}

\author{B. Andrei Bernevig}
\email{bernevig@princeton.edu}
\affiliation{Department of Physics, Princeton University, Princeton, New Jersey 08544, USA}
\affiliation{Donostia International Physics Center, P. Manuel de Lardizabal 4, 20018 Donostia-San Sebastian, Spain}
\affiliation{IKERBASQUE, Basque Foundation for Science, Bilbao, Spain}

\begin{abstract}
Topological band theory has achieved great success in the high-throughput search for topological band structures both in paramagnetic and magnetic crystal materials. However, a significant proportion of materials are topologically trivial insulators at the Fermi level. 
In this paper, we show that, remarkably, for a subset of the topologically trivial insulators, knowing \emph{only} their electron number and the Wyckoff positions of the atoms we can separate them into two groups: the obstructed atomic insulator (OAI) and the atomic insulator (AI).
The interesting group, the OAI, have a center of charge not localized on the atoms. 
Using the theory of topological quantum chemistry, in this work we first derive the necessary and sufficient conditions for a topologically trivial insulator to be a filling enforced obstructed atomic insulator (feOAI) in the 1651 Shubnikov space groups. 
Remarkably, the filling enforced criteria enable the identification of obstructed atomic bands without knowing the representations of the band structures.
Hence, no ab-initio calculations are needed for the filling enforced criteria, although they are needed to obtain the band gaps.
With the help of the \webTQC, we have performed a high-throughput search for feOAIs and have found that \TQCDBNbrICSDsFeOAI\ ICSD entries (\TQCDBNbrMaterialsFeOAI\ unique materials) are paramagnetic feOAIs, among which \TQCDBNbrICSDsFeOAIIndirectGap\ (\TQCDBNbrMaterialsFeOAIIndirectGap) materials have an indirect gap. 
The metallic obstructed surface states of feOAIs are also showcased by several material examples. 

\end{abstract}

\maketitle

\addtocontents{toc}{\protect\setcounter{tocdepth}{0}}
\addtocontents{lot}{\protect\setcounter{lotdepth}{-1}}

\section{Introduction}
The development of topological quantum chemistry (TQC) \cite{QuantumChemistry,MTQC,Chen-MTQC} and equivalent methods of symmetry indicator \cite{AshvinIndicators,AshvinMagnetic} have led to high-throughput discoveries of topological quantum materials \cite{tqcmaterials,ChenMaterials,AshvinMaterials,xu2020high,Vergniory2021}. (See \webTQC\ for examples.)
In the theory of TQC \cite{QuantumChemistry}, topologically trivial insulators are defined as band representations (BRs), which are equivalent to a set of symmetric exponentially decayed Wannier functions. 
Any BR can be spanned by the elementary BRs (EBRs), which are induced from irreducible representations (irreps) at maximal Wyckoff positions (WPs)  \cite{QuantumChemistry,MTQC,Bandrep1,Bandrep2,AlexeyVDBWannier}. 
For a given material, a set of isolated bands below the Fermi level can be characterized by an integer vector, the symmetry-data-vector, whose components are the multiplicities of irreps formed by the bands at the high symmetry  momenta \cite{PhysRevB.97.035139,Bandrep1,Bandrep2,cano2020band}. 
If the symmetry-data-vector can be expressed as a non-negative integer linear combination of the symmetry-data-vectors of EBRs, the set of bands is eigenvalue-diagnosed as trivial by TQC, which are referred as ``TQC trivial insulator".
Otherwise the material must be topological. 
If some of the integer coefficients  in the decomposition of the symmetry-data-vector in terms of EBRs must necessarily be negative, the material is identified as fragile topological \cite{QuantumChemistry,AshvinFragile,BarryFragile,ZhidaFragile2,song2020fragile,FragileFlowMeta,MTQC,KooiPartialNestedBerry,AdrianFragile,KoreanFragile,PhysRevResearch.1.032005,ArisFragile}. If some of the coefficients must necessarily be  rational fractions, the material is a stable topological insulator  \cite{CharlieTI,AndreiTI,kitaev2009periodic,HOTIBernevig,QuantumChemistry,AshvinIndicators,HOTIBismuth,MTQC,ChenTCI,AshvinTCI,Chen-MTQC} or topological semimetal \cite{ZhidaSemimetals,MTQC,Chen-MTQC}. 
Finally, if the irreps of the bands below the Fermi level break the so-called compatibility relations, the material is an enforced semimetal with band crossings in high symmetry lines or high symmetry planes \cite{QuantumChemistry,Bandrep1,AshvinIndicators}. 

\begin{figure}
\centering\includegraphics[width=3.2in]{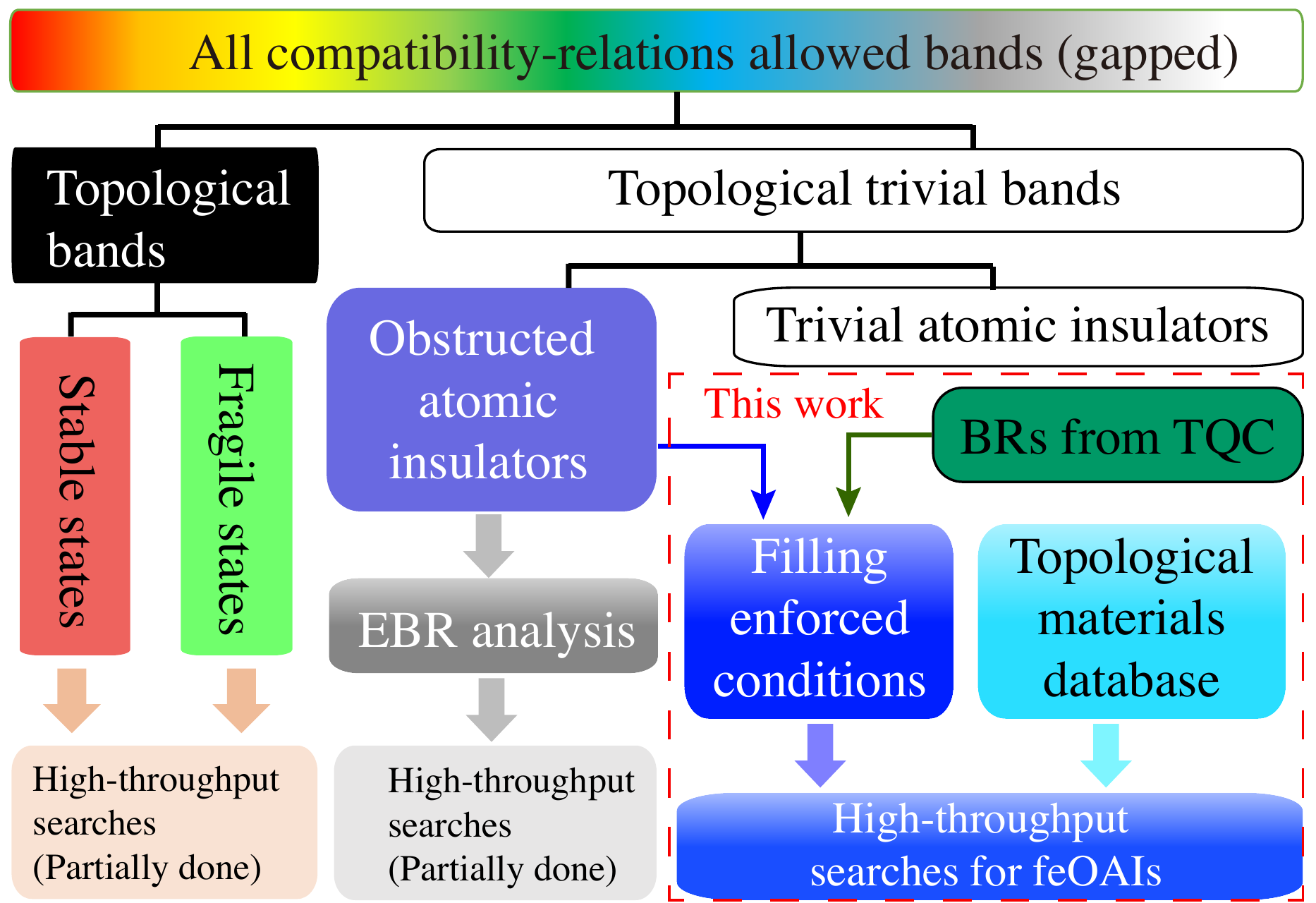}
\caption{Topological classification of the gapped bands satisfying compatibility relations. High-throughput searches of topological bands, including strong topological bands and fragile bands, are partially done. Topologically trivial bands can be classified into obstructed atomic limit bands and trivial atomic bands. Although a full topological classification of obstructed bands has not been achieved, we perform the first high-throughput search of filling enforced obstructed bands in the present work.}\label{fig1}
\end{figure}

As schematically shown in Figure\ref{fig1}, 
several high-throughput calculations \cite{tqcmaterials,ChenMaterials,AshvinMaterials,xu2020high} have identified thousands of compounds that belong to different topological categories. 
However, among the topologically trivial insulating materials, there are still interesting cases in which the valence bands are BRs (in terms of irreps) but not BRs induced from the \emph{occupied} WPs.
In other words, these insulators are not atomic insulators where electrons fill atomic orbitals at the atom sites. 
We refer to such materials as the obstructed atomic insulators (OAI) \cite{QuantumChemistry,PhysRevB.97.035139,cano2020band,QueirozOAL,AshvinFragile,Soluyanov11,ArisInversion}.
In the decomposition of the symmetry-data-vector an OAI, there must be at least one BR induced from an empty Wyckoff position. 
We refer to this empty site as an obstructed Wannier charge center (OWCC).
Given an OAI with the cleavage terminations cutting through an OWCC, which is off the occupied sites, there has to exist metallic surface states in the gap between valence and conduction bands, which is referred as \emph{filling anomaly} \cite{multipole,WladTheory,PhysRevResearch.1.033074}.

\begin{figure}
\centering\includegraphics[width=3.2in]{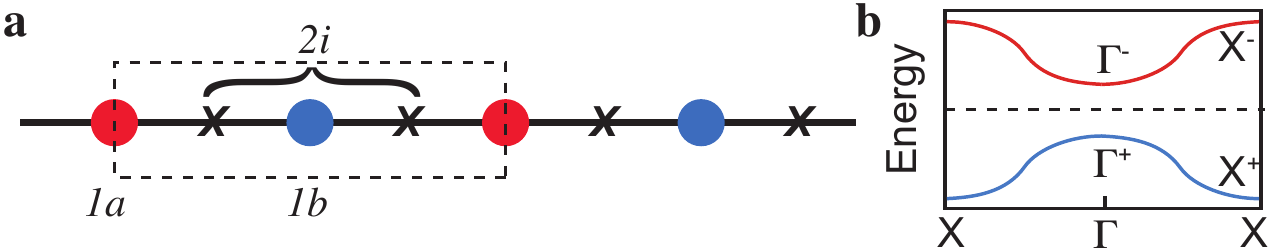}
\caption{FeOAI in 1-d. (a) A 1-d atomic chain with a Wyckoff position $2i$ occupied by atoms. (b) Schematic diagram of the band structure of the 1-d model in (a). The blue and red bands are valence and conduction bands, respectively.}\label{fig2}
\end{figure}

In general, the OAI can be detected from the identified irreps at all the high symmetry momenta through the RSI \cite{ZhidaFragile2}. 
In some special cases, it is possible to identify the OAIs just by electron counting \emph{without} ab-initio calculations.
In some space groups (SGs) and magnetic space groups (Shubnikov space groups, in general \cite{ShubnikovBook}), knowing the number of valence electrons and the occupied Wyckoff positions (WPs) is sufficient for identifying topologically trivial materials as OAIs. 
We denote this special type of OAI as filling enforced obstructed atomic insulator (feOAI). Figure~\ref{fig2}(a) shows a simple example of a feOAI. It represents a Su-Schrieffer-Heeger (SSH)
chains model \cite{PhysRevLett.42.1698} in the space group $P\bar{1}$ with $2i$ sites occupied by atoms. From TQC, a necessary condition for this atomic chain to be a band insulator is that the number of electrons $N_e$ in a unit cell  is even, i.e. $N_e=2n (n=1,2,3,...)$. As the dimension of the BR induced from single orbitals at $2i$ is 4, when the number of electrons is an odd multiple of 2, i.e. $N_e=4n+2 (n=0,1,2,...)$, the decomposition of the BR into EBRs needs at least one EBR induced from an empty WP, $1a$ or $1b$, because these EBRs have dimension 2. 
Thus the filling enforced condition $N_e=4n+2 (n=0,1,2,...)$ plus the XRD information that the atoms are place at the $2i$ Wyckoff position alone can identify whether an insulator is feOAI or not without any information of the wavefunction at the high symmetry points. To obtain the OWCC of feOAI (in our example, to elucidate whether the needed EBR or EBRs out of WP $2i$ are those of $2a$ or $2b$), one needs to analyze the specific BR. 
As schematically shown in Figure~\ref{fig2}(b), when the irreps of the valence band are $\{\Gamma^+ X^+\}$, which are induced from $A_g @ 1a$, the OWCC is sitting at $1a$.

In this paper we first derive in Section~\ref{section: FOC} the filling enforced conditions of topologically trivial bands to be obstructed for all the 1651 Shubnikov space groups (SSGs). In Section~\ref{section: HTS}, using the obtained filling enforced conditions, we perform the first high-throughput search for feOAIs from the \webTQC. In Section~\ref{section:SS} we analyze some of their surface states. 

\section{Filling Enforced Conditions}
\label{section: FOC}

In this section we will obtain the filling enforced conditions of OAIs in all the 1651 SSGs, which contain as a particular case the SGs of non-magnetic structures, both with and without spin-orbit coupling (SOC).

Let's consider an insulator that has been identified as trivial using the TQC \cite{QuantumChemistry} or the magnetic TQC \cite{MTQC} method (it has been labeled as LCEBR in \cite{tqcmaterials} or \cite{xu2020high}) in a single (double) SG with a set of WPs $\{\alpha\}$ with multiplicities $\{n_{\alpha}\}$. We consider here the multiplicities in a primitive unit cell (not the conventional cell used in the International Tables of Crystallography \cite{ITCA}). 
We denote the allowed co-representations of the site-symmetry group of the WP $\{\alpha\}$ as $\{\rho_{\alpha}^i\}$, whose dimensions are $\{d(\rho_{\alpha}^i)\}$, with $i=1\ldots N_{rep,\alpha}$, where $N_{rep,\alpha}$ is the total number of co-representations of the site-symmetry group $G_{\alpha}$. In paramagnetic compounds, for which time-reversal symmetry is always a symmetry operation of the SG, if SOC is considered (spinful cases), the Kramers theorem implies that the dimension $\{d(\rho_{\alpha}^i)\}$ is always an even number. In paramagnetic spinless (without SOC) systems the dimensions of the irreps can be odd, but two times of the dimension electrons are needed to fill each irrep due to the 2-fold spin degeneracy. 
Finally, in magnetic groups, where the time-reversal and spin SU(2) symmetries are absent, there is no restriction on electron number parity as in the paramagnetic groups. 
All the band co-representations $\{\rho_{\alpha}^i\}$ (and the corresponding dimensions $\{d(\rho_{\alpha}^i)\}$) induced from any WP in the 1651 double SSGs are accessible on the Bilbao Crystallographic Server through the \href{http://www.cryst.ehu.es/cryst/mbandrep}{MBANDREP} program \cite{MTQC}. The complete set of elementary band co-representations are also listed in this program. The subset of co-representations of magnetic type-I and type-II (the last ones are the relevant groups for paramagnetic phases) and the elementary band co-representations are also implemented in the \href{http://www.cryst.ehu.es/cryst/mbandrep}{BANDREP} program \cite{Bandrep1}.

Now we assume that there are $M$ occupied WPs $\alpha_1 \ldots \alpha_M$. Some of these WPs can share the same label because they can belong to (sometimes maximal) WPs  whose coordinates depend on continuous parameters, and the values of these parameters are different. If an atom sits  at a given WP $\alpha_j$ and contributes $N_{j}$ electrons (typically these are the electrons in the outer shell of the atom), the whole WP $\alpha_j$ contributes $n_{\alpha_j}N_{j}$ electrons to the total number of electrons in a primitive unit cell, with $n_{\alpha_j}$ being the multiplicity (in each primitive unit cell) of the Wyckoff position $\alpha_j$.
Then, the total number of spinless (spinful) electrons per primitive cell in the system can be expressed as, 
\begin{equation}
N_e= \sum_{j=1}^M n_{\alpha_j} N_{j}
\label{eq1}
\end{equation}
A necessary (but not sufficient) condition for a material to be \textit{eigenvalue indicated topologically trivial insulator} is that there exists a solution to the equation:
\begin{eqnarray}
N_e=\sum_{j=1}^{\text{All WPs}}  \sum_{i=1}^{N_{rep, \alpha_j}} n_{\alpha_j}  d(\rho^i_{\alpha_j}) N_{i,j}
\label{ti}
\end{eqnarray} 
where $j$ sums over all the WPs of the corresponding SSG, and $N_{i,j}$ are non-negative integers that represent how many ``orbitals" (representations $\rho^i_{\alpha_j}$) at WP $\alpha_j$ are occupied. Due to the spin degeneracy when SOC is neglected, each orbital $\rho^i_{\alpha_j}$ can be occupied by two electrons. In this case, we should multiply the right side of Eq.~(\ref{ti}) by 2. However, in order to use a single expression for the spinful and spinless cases, in Eq.~(\ref{ti}) and in the rest of this paper, $N_e$ will be equal to the number of electrons when the SOC is considered, and to one half the number of electrons in the absence of SOC. With this convention Eq.~(\ref{ti}) is valid in both cases.

For each (double) SSG, the dimensions of the BRs induced from the co-irreps $\rho_{\alpha}^i$ of the site symmetry group of each WP $\alpha$ can be expressed as,

\begin{equation}
d(\rho_{\alpha}^i\uparrow\mathcal{G})=n_{\alpha}d(\rho_{\alpha}^i)
\label{eq3}
\end{equation} 
Then Eq.~(\ref{ti}) can be rewritten as,

	\begin{equation}
	 N_e=\sum_{j=1}^{\text{All WPs}}  \sum_{i=1}^{N_{rep, \alpha_j}} d(\rho_{\alpha_j}^i\uparrow\mathcal{G}) N_{i,j}, N_{i,j} \ge 0, \in \mathbf{Z} \label{eq4}
	\end{equation}

For symmetry eigenvalue indicated trivial insulators, the solution for Eq.~(\ref{eq4}) is in general not unique. 
The set of EBRs in a SSG form an overcomplete basis of the BRs in a group and, moreover, in our analysis we must consider all the WPs of the SSG, maximal and non-maximal. It is well known (\cite{ZakBandrep2,ZakException2}) that the band co-representations induced from non-maximal WPs are linear combinations of EBRs (induced from maximal WPs); therefore different sets of $N_{i,j}$ coefficients can in principle satisfy Eq.~(\ref{eq4}).
  
However, if the first summation in Eq.~(\ref{eq4}) is restricted only to the $M$ \emph{atomic-occupied} WPs and not to the whole set of WPs of the SSG, the resulting equation,
	\begin{eqnarray}
N_e=\sum_{j=1}^{M}  \sum_{i=1}^{N_{rep, \alpha_j}} d(\rho_{\alpha_j}^i\uparrow\mathcal{G}) N_{i,j}, \;\;\; N_{i,j} \ge 0, \in \mathbf{Z}
	\label{tti}
	\end{eqnarray} 

may have less solutions or no solution.
Absence of solution to Eq.~(\ref{tti}) is a sufficient condition for OAI.

To summarize, a material identified as trivial insulator by TQC always satisfies Eq.~(\ref{eq4}), but if Eq.~(\ref{tti}) is not satisfied, mathematically
  \begin{equation}
  \nexists N_{i,j}\ge0, \in \mathbf{N}, s.t. N_e=\sum_{j=1}^{M}\sum_{i=1}^{N_{rep,\alpha_j}}d(\rho_{\alpha_j}^i\uparrow\mathcal{G})N_{i,j}
  \label{eq9}  
  \end{equation}
then  the material is a feOAI. Although it may be possible to define a set of localized Wannier functions, at least one of them must be centered at an unoccupied WP.

In the Appendices \ref{230SG} and \ref{230DSG} we have tabulated the dimensions of all the BRs for all the 230 space groups (SGs) and double space groups (DSGs), respectively. In the Appendices \ref{dimTypeINoSOC}-\ref{dimTypeIVSOC}, we have tabulated the dimensions of all the BRs for all the single and double magnetic groups (Shubnikov groups of types I, II and IV). Once the occupied WPs in a material and the number of electrons in the primitive unit cell are known, these tables together with Eqs.~(\ref{eq4}) and (\ref{eq9}) can be used to check whether a material that has been identified by TQC as a topologically trivial insulator is a feOAI. In Section~\ref{fec_DSG} of the Appendix, we have tabulated the feOAI conditions for all the 230 space groups. The conditions are exactly the same for single (without SOC) and double (with SOC) space groups. In sections \ref{fecTypeINoSOC}-\ref{fecTypeIVSOC} of the Appendix we have also tabulated the feOAI conditions for all the 1651 SSGs. Unlike for non-magnetic systems, the conditions are different for single and double SSGs (More discussion about this is detailed in Appendix~\ref{fec_1651SSG}).

We will apply the above analysis to a simple example. In the DSG $Pbca$ (\# 61) the multiplicities of the Wyckoff positions $a,b$ and $c$ are 4,4 and 8, respectively, and the allowed dimensions of the BRs induced from the irreps (with SOC) at $a,b$ and $c$  are $8,8$ and $16$, respectively (see Table~\ref{230DSG}). If the material has been identified as trivial insulator using TQC and the $a$ and/or $b$ WPs are occupied, the number of valence electrons has to be a multiple of 8, and the Wannier centers can be located at the occupied Wyckoff position(s). However, if only WP $c$ is occupied by atoms and the number of electrons is $N_e=8+16\mathbb{N}$ (with $\mathbb{N}$ being integers), the Wannier functions cannot be all located at the occupied WPs because otherwise the total electron number would be a multiple of 16. 
We emphasize that the filling $N_e=8+16\mathbb{N}$ can be realized if, for example, a single $c$ position is occupied by atoms and each atom at $c$ contributes an odd number of electrons.
At least one of them must be centered at $a$ or $b$. The condition given by Eq.~(\ref{tti}) cannot be fulfilled by any set of non-negative integers $N_{i,j}$ and the material is thus a feOAI.

Equations (\ref{eq4}) and (\ref{eq9}) give the necessary and sufficient conditions for a material to be a feOAI once the TQC method has identified it as a topologically trivial insulator.  Alternatively, we can  establish  sufficient but not necessary  conditions to have a feOAI material that are easier to use than Eq.~(\ref{eq9}). Concretely, if the following two conditions are simultaneously fulfilled, Eqs.~(\ref{eq4}) and (\ref{eq9}) are also satisfied and the material is a feOAI: 
\begin{eqnarray}
N_e/gcd_{\textrm{all}}(\{d(\rho_{\alpha_j}^i\uparrow\mathcal{G})\}) \in \mathbf{Z} 
\label{eq5}  \\
N_e/gcd_{\textrm{occ}}(\{d(\rho_{\alpha_j}^i\uparrow\mathcal{G})\}) \notin \mathbf{Z} \Longrightarrow \text{feOAI}  
\label{eq6}  
\end{eqnarray} 
Here $gcd$ stands for ``greatest common denominator" and the subindices ``all'' in Eq.~(\ref{eq5}) and ``occ''  in Eq.~(\ref{eq6}) mean that the $gcd$ must be calculated using the BRs of all the WPs of the group and only the occupied WPs, respectively. Eq.~(\ref{eq5}) guarantees that the material can be an insulator and Eq.~(\ref{eq6}) implies  that the insulator is an feOAI.
Our previous example based on SG $Pbca$ (\# 61) fulfills these two conditions with $N_e=8$, $gcd_{\textrm{all}}(\{d(\rho_{\alpha_j}^i\uparrow\mathcal{G})\})=8$ and $gcd_{\textrm{occ}}(\{d(\rho_{\alpha_j}^i\uparrow\mathcal{G})\})=16$.

However, as a counter-example that demonstrates that Eqs.~(\ref{eq5}) and (\ref{eq6}) are not necessary conditions, let's consider the DSG $P\overline{3}1m$ (\# 162) with SOC and  occupied WPs $3f$ (or $3g$) and $4h$ with multiplicities 3 and 4, respectively, and BRs of dimension 6 and 8, respectively (see Table~\ref{230DSG}). In this DSG  $gcd_{\textrm{all}}(\{d(\rho_{\alpha_j}^i\uparrow\mathcal{G})\})=gcd_{\textrm{occ}}(\{d(\rho_{\alpha_j}^i\uparrow\mathcal{G})\})=2$. If $N_e=10$, which is possible when both $3f$, $3g$ and a single $4h$ WP are occupied by atoms with a single electron in the outer shell, the condition (\ref{eq5}) is satisfied but (\ref{eq6}) not. However, the insulator is a feOAI, because it is impossible to fill in complete bands induced from WPs $3f/3g$ and $4h$, whose dimensions are 6 and 8, respectively, to get as a result a BR of dimension 10. This shows that to avoid missing feOAIs one should solve the more involved Eq.~(\ref{eq9}). (See Appendix \ref{fec_DSG} for all the SGs of the same case).

In general, Eq.~(\ref{eq9}) is analytically unsolvable and is related to the Frobenius coin problem \cite{alfonsin2005diophantine}. However, for our particular case, since the number of distinct dimensions of BRs in all the SSG is small, the Frobenius coin problem is solvable by exhaustion. We have determined all the solutions to Eq.~(\ref{eq4}) and (\ref{eq9}) in the 1651 SSGs with and without SOC. In Appendix \ref{fec_DSG} we list the feOAI conditions for the 230 SGs (type-II SSGs), relevant in the analysis of non-magnetic structures. The table gives the conditions for spinful and spinless irreps, which are exactly the same once we consider that in the spinless case $N_e$ represents half the number of electrons. In Appendix \ref{fec_1651SSG} we have included the feOAI conditions for the magnetic space groups (Shubnikov groups of type I, III and IV) with and without SOC. For magnetic groups, the solutions to Eq.~(\ref{eq4}) and (\ref{eq9}) with and without SOC are different, in general, and they are given in separate tables. Note that in the absence of SOC, the spinless feOAI conditions are only applied to the antiferromagnetic systems which are independent of the spin but not the ferromagnetic ones.

\section{High-throughput search for feOAIs}
\label{section: HTS}

By application of the filling enforced conditions obtained in section III to all the paramagnetic trivial insulators in the \webTQC, we have performed the first high-throughput search of paramagnetic feOAIs. 
Among the \TQCDBNbrICSDsTrivial\ ICSD entries (\TQCDBNbrMaterialsTrivial\ {\it unique materials}, i.e., ICSDs sharing the same stoichiometric formula, space group and topological property at the Fermi energy) that are identified as trivial insulators in the database, we have found \TQCDBNbrICSDsFeOAI\ ICSD entries (\TQCDBNbrMaterialsFeOAI\ unique materials), that satisfy the filling enforced conditions and hence they are feOAIs. 
This represents about 3\% of the total number of topologically trivial insulators. In Table~\ref{feoai_count}, we tabulate the numbers of feOAIs found in the high-throughput search for each DSG (Type-II SSG with SOC). 
Among the \TQCDBNbrICSDsFeOAI\ ICSD entries (\TQCDBNbrMaterialsFeOAI\ unique materials) identified as feOAIs, \TQCDBNbrICSDsFeOAIIndirectGap\ (\TQCDBNbrMaterialsFeOAIIndirectGap) materials have a finite indirect band gap in the whole BZ. 
The full list of materials, with detailed electronic properties, are tabulated in Appendix \ref{feoai_list}. 

Using the filling enforced conditions of MSGs, we have also found one magnetic feOAIs candidate from the 403 magnetic materials on the \webMTQC. 
We find that the topologically trivial phase of Er$_2$Ni$_2$In with \bcsidweblong{1.195} satisfies the filling enforced condition of the Type-IV MSG $C_amcm$ (\#63.467). Hence, it is a magnetic feOAI. See Appendix \ref{mag_feoais} for more details of Er$_2$Ni$_2$In.

\begin{table*}
\centering
\begin{tabular}{>{\columncolor[gray]{0.8}}l|c||>{\columncolor[gray]{0.8}}l|c||>{\columncolor[gray]{0.8}}l|c||>{\columncolor[gray]{0.8}}l|c||>{\columncolor[gray]{0.8}}l|c||>{\columncolor[gray]{0.8}}l|c||>{\columncolor[gray]{0.8}}l|c||>{\columncolor[gray]{0.8}}l|c||>{\columncolor[gray]{0.8}}l|c}
\hline\hline
 SG & Count & SG & Count & SG & Count & SG & Count & SG & Count & SG & Count & SG & Count & SG & Count & SG & Count \\
\hline
2 & \TQCDBNbrFEOAIPerSG{2} &5 & \TQCDBNbrFEOAIPerSG{5} &12 & \TQCDBNbrFEOAIPerSG{12} &13 & \TQCDBNbrFEOAIPerSG{13} &14 & \TQCDBNbrFEOAIPerSG{14} &15 & \TQCDBNbrFEOAIPerSG{15} &20 & \TQCDBNbrFEOAIPerSG{20} &38 & \TQCDBNbrFEOAIPerSG{38} &41 & \TQCDBNbrFEOAIPerSG{41}  \\
\hline
43 & \TQCDBNbrFEOAIPerSG{43} &45 & \TQCDBNbrFEOAIPerSG{45} &55 & \TQCDBNbrFEOAIPerSG{55} &58 & \TQCDBNbrFEOAIPerSG{58} &60 & \TQCDBNbrFEOAIPerSG{60} &61 & \TQCDBNbrFEOAIPerSG{61} &63 & \TQCDBNbrFEOAIPerSG{63} &64 & \TQCDBNbrFEOAIPerSG{64} &65 & \TQCDBNbrFEOAIPerSG{65}  \\
\hline
69 & \TQCDBNbrFEOAIPerSG{69} &70 & \TQCDBNbrFEOAIPerSG{70} &71 & \TQCDBNbrFEOAIPerSG{71} &81 & \TQCDBNbrFEOAIPerSG{81} &82 & \TQCDBNbrFEOAIPerSG{82} &86 & \TQCDBNbrFEOAIPerSG{86} &87 & \TQCDBNbrFEOAIPerSG{87} &88 & \TQCDBNbrFEOAIPerSG{88} &91 & \TQCDBNbrFEOAIPerSG{91}  \\
\hline
92 & \TQCDBNbrFEOAIPerSG{92} &102 & \TQCDBNbrFEOAIPerSG{102} &114 & \TQCDBNbrFEOAIPerSG{114} &118 & \TQCDBNbrFEOAIPerSG{118} &122 & \TQCDBNbrFEOAIPerSG{122} &126 & \TQCDBNbrFEOAIPerSG{126} &136 & \TQCDBNbrFEOAIPerSG{136} &137 & \TQCDBNbrFEOAIPerSG{137} &138 & \TQCDBNbrFEOAIPerSG{138}  \\
\hline
139 & \TQCDBNbrFEOAIPerSG{139} &141 & \TQCDBNbrFEOAIPerSG{141} &142 & \TQCDBNbrFEOAIPerSG{142} &147 & \TQCDBNbrFEOAIPerSG{147} &148 & \TQCDBNbrFEOAIPerSG{148} &152 & \TQCDBNbrFEOAIPerSG{152} &162 & \TQCDBNbrFEOAIPerSG{162} &164 & \TQCDBNbrFEOAIPerSG{164} &165 & \TQCDBNbrFEOAIPerSG{165}  \\
\hline
166 & \TQCDBNbrFEOAIPerSG{166} &167 & \TQCDBNbrFEOAIPerSG{167} &194 & \TQCDBNbrFEOAIPerSG{194} &197 & \TQCDBNbrFEOAIPerSG{197} &199 & \TQCDBNbrFEOAIPerSG{199} &202 & \TQCDBNbrFEOAIPerSG{202} &204 & \TQCDBNbrFEOAIPerSG{204} &205 & \TQCDBNbrFEOAIPerSG{205} &214 & \TQCDBNbrFEOAIPerSG{214}  \\
\hline
215 & \TQCDBNbrFEOAIPerSG{215} &216 & \TQCDBNbrFEOAIPerSG{216} &217 & \TQCDBNbrFEOAIPerSG{217} &220 & \TQCDBNbrFEOAIPerSG{220} &225 & \TQCDBNbrFEOAIPerSG{225} &227 & \TQCDBNbrFEOAIPerSG{227} & \bf Total & \bf \TQCDBNbrFEOAIPerSG{0} & & & \\
\hline\hline
\end{tabular}
\caption{Statistic of the numbers of paramagnetic feOAIs in each DSG. For each DSG, we provide the number of ICSD entries that are paramagnetic feOAIs, and the number of unique materials, i.e., ICSDs sharing the same stoichiometric formula, space group and topological property at the Fermi energy, in parentheses.}\label{feoai_count}
\end{table*}

\section{Surface states of feOAIs}
\label{section:SS}

The occupied states of symmetry-protected topological insulators are delocalized, and these systems present gapless states on lower dimensional  edges/surfaces. In contrast, the occupied states of OAIs are localized.
However, as all the OWCCs are  localized away from  the atom sites, it is still possible to have a gapless surface/edge as long as the cleavage cuts through the OWCCs. 
In this section, we take CdSb  and its metallic surface states due to the OWCC as a showcased feOAI.

\begin{figure}[htbp]
\centering\includegraphics[width=3.2in]{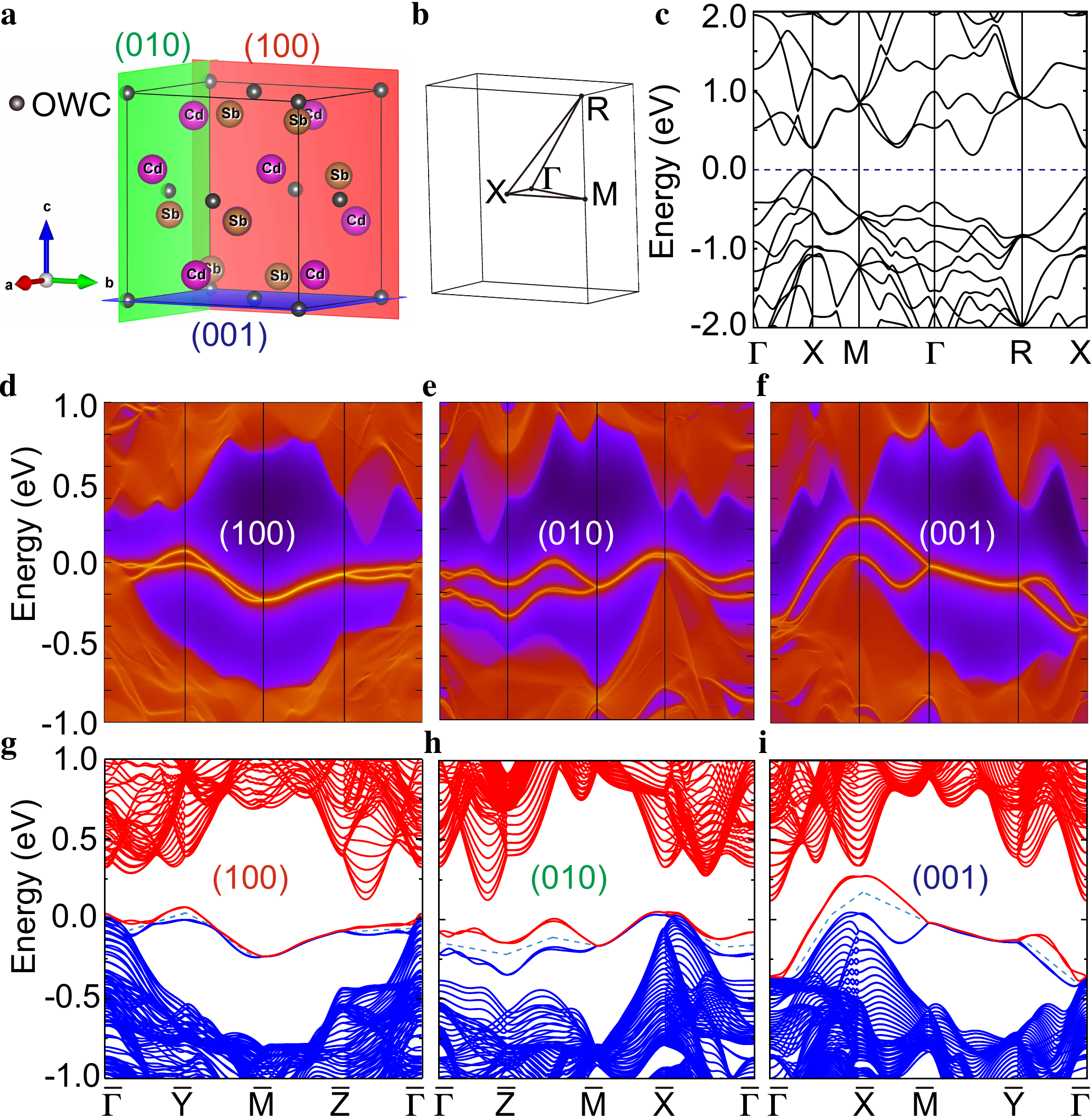}
\caption{(a) Crystal structure of CdSb, where the black spheres are the position of OWCCs. 
The red, green and blue planes cutting through the OWCCs are the cleavage planes with Miller indices (100), (010) and (001), respectively. (b) The 3D Brillouin zone (BZ) for CdSb. (c) Electronic band structure  for CdSb with SOC along the high-symmetry paths in BZ.
(d)-(f) Surface states of semi-infinite CdSb on the cleavage planes defined in (a). The surface states are highlighted by bright orange lines in the gap. (g)-(i) Surface states of CdSb with slab structure along the cleavage planes defined in (a). By counting the number of bands from the lowest bands until charge neutrality, we indicate the valence and conduction bands by blue and red lines, respectively. From these figures one can infer that the surface states in (d)-(f) are half-filled.
}\label{fig3}
\end{figure}

As shown in Figure~\ref{fig3}(a), CdSb adopts an orthorhombic lattice with space group  $Pbca$ (\# 61) \cite{range1988cadmium}, whose generators can be taken as the inversion $\{I|000\}$ together with the two two-fold screw rotations $\tilde{C}_{2y}=\{C_{2y}|(0,1/2,1/2)\}$ and $\tilde{C}_{2z}=\{C_{2z}|(1/2,0,1/2)\}$. Its lattice parameters are $a=6.469\text{\AA},b=8.251\text{\AA}$
and $c=8.522\text{\AA}$. 
In CdSb both Cd and Sb are located at the general Wyckoff position $8c(x,y,z)$ with 
coordinates  $(0.5503,0.6238,0.63426)$ and $(0.1398,0.0739,0.1039)$ respectively.
Figure~\ref{fig3}(c) shows the band structure of this compound, that has been identified as a topologically trivial insulator with an indirect gap of $\sim0.2 eV$ (see \webTQC $ $  or \href{http://materiae.iphy.ac.cn}{Materiae} with \icsdweb{52830} \cite{ICSD}, for instance).

The outer-shell electronic configurations of Cd and Sb are $4d^{10}5s^2$ and $5s^25p^3$, respectively.  As both Cd and Sb sit at a Wyckoff position of multiplicity $8$, the total number of spinful electrons in a primitive unit cell is $N_e=56$ (if we assume $5s^2$ for Cd), which is divisible by $8$ but not by $16$. As discussed in Section~\ref{section: HTS}, CdSb satisfies the filling enforced conditions for SG $Pbca$, i.e. $N_e \in 8N$ and $N_e \notin 16N$. So, CdSb is a feOAI, without the need for any ab-initio calculations.
 
From \webTQC, the irreps of CdSb at maximal, high-symmetry $k$ points can be decomposed as 
linear combination of EBRs (LCEBR). All possible decompositions are shown in Table~\ref{table3}. 
For each decomposition, there are always orbitals whose BR is not inducible pinned at $4a$.
This signals the existence of obstructed orbitals centered at $4a$ 
and any cleavage plane cutting through the $4a$ position has metallic surface states at the Fermi level. Based on the first-principle calculations performed on the Vienna {\it ab-initio} Simulation Package (VASP) \cite{VASP1996}, we have constructed the maximally localized Wannier function \cite{Souza2001Maximally-PRB} of CdSb and calculate its surface states \cite{WU2017} along different crystal directions.
The calculated surface states of CdSb on the (100), (010) and (001) surfaces are shown in Figure~\ref{fig3}(d)-(f). Counting the number of valence bands shows  that the surface states indicated by red lines near the Fermi level are half filled. The BR of the obstructed surface states on each surface is one EBR of the plane subgroup of DSG $Pbca$ on the corresponding 2D surface. All  these EBRs of the 3c surface are indecomposable and hence the surface states on each surface have to be metallic, which is also referred to as a filling anomaly.

In Appendix \ref{ss_calculations}, we have also calculated the obstructed surface states of additional four feOAIs, including GaSe, GaS, IrP$_2$ and CuP$_2$.

\begin{table}
\centering
\begin{tabular}{c|c|c|c|c}
\hline\hline
case & $\overline{A_gA_g}@4a$ &$\overline{A_uA_u}@4a$ & $\overline{A_gA_g}@4b$ & $\overline{A_uA_u}@4b$\tabularnewline
\hline
 \#1 & 9 & 8 & 0 & 0\\
\hline
\#2 & 8 & 7 & 1 & 1\\
\hline
\#3 & 7 & 6 & 2 & 2\\
\hline
\#4 & 6 & 5 & 3 & 3\\
\hline
\#5 & 5 & 4 & 4 & 4\\
\hline
\#6 & 4 & 3 & 5 & 5\\
\hline
\#7 & 3 & 2 & 6 & 6\\
\hline
\#8 & 2 & 1 & 7 & 7\\
\hline
\#9 & 1 & 0 & 8 & 8\\
\hline\hline
\end{tabular}
\caption{All possible decompositions of the BR of CdSb into linear combination of the EBRs in DSG $Pbca$ (\# 61). The first column gives the EBRs induced from different orbitals at Wyckoff positions $4a$ or $4b$; the numbers below are the multiplicities of each EBR in the corresponding decomposition.}\label{table3}
\end{table}

\section{Conclusions}
In this work we have derived, for all the 1651 SSGs, the filling enforced conditions that guarantee that a material identified as a topologically trivial band insulator is in the obstructed atomic insulating phase. These conditions provide an efficient ab-initio free way to search for paramagnetic and magnetic materials in the obstructed atomic phase. 
Through the application of the filling enforced conditions to the non-magnetic structures labeled as trivial insulators in the \webTQC, we remarkably find \TQCDBNbrICSDsFeOAI\ ICSD entries (\TQCDBNbrMaterialsFeOAI\  unique materials) as feOAIs, among which 750 (\TQCDBNbrMaterialsFeOAIIndirectGap) compounds have an indirect band gap. Combined with the BR analysis and first principles calculations, we have also showcased the filling anomaly metallic surface states for specific surfaces of several feOAIs. The special metallic surface states in feOAIs provide an ideal platform for the study of two-dimensional electron gases that could be detected in ARPES or STM experiments.

{\bf Acknowledgements}
We acknowledge the computational resources Cobra/Draco in the Max Planck Computing and Data Facility (MPCDF).
This work is part of a project that has received funding from the European Research Council (ERC) under the European Union's Horizon 2020 research and innovation programme (grant agreement no. 101020833. B.A.B., N.R. and Z-D.S. were also supported by the U.S. Department of Energy (Grant No. DE-SC0016239), and were partially supported by the National Science Foundation (EAGER Grant No. DMR 1643312), a Simons Investigator grant (No. 404513), the Office of Naval Research (ONR Grant No. N00014-20-1-2303), the Packard Foundation, the Schmidt Fund for Innovative Research, the BSF Israel US foundation (Grant No. 2018226), the Gordon and Betty Moore Foundation through Grant No. GBMF8685 towards the Princeton theory program, a Guggenheim Fellowship from the John Simon Guggenheim Memorial Foundation and the NSF-MRSEC (Grant No. DMR-2011750). B.A.B. and N.R. gratefully acknowledge financial support from the Schmidt DataX Fund at Princeton University made possible through a major gift from the Schmidt Futures Foundation. L.E. was supported by the Government of the Basque Country (Project IT1301-19) and the Spanish Ministry of Science and Innovation (PID2019-106644GB-I00). 
M.G.V. thanks support from the Spanish Ministry of Science and Innovation  (grant  number  PID2019-109905GB-C21).
C.F. was supported by the European Research Council (ERC) Advanced Grant No.  742068 ``TOP-MAT'', Deutsche Forschungsgemeinschaft (DFG) through SFB 1143, and the Würzburg-Dresden Cluster of Excellence on Complexity and Topology in Quantum Matter-ct.qmat (EXC 2147, Project No. 390858490). S.S.P.P. acknowledges funding by the Deutsche Forschungsgemeinschaft (DFG, German Research Foundation) – Project number 314790414.
J.L.M. has been supported by Spanish Science Ministry grant PGC2018-094626-B-C21 (MCIU/AEI/FEDER, EU) and by Basque Government grant IT979-16. L.E., M.G.V. and N.R. were also sponsored by a Global Collaborative Network grant. Y.X. received support from the Max Planck Society.

\bibliography{ref}

\clearpage

\onecolumngrid
\renewcommand{\thesection}{Appendix \arabic{section}}
\renewcommand{\thesubsection}{\arabic{section}.\arabic{subsection}}
 
\appendix 
\clearpage
\begin{center}
{\bf Supplementary materials of "Filling-Enforced Obstructed Atomic Insulators"}
\end{center}

\tableofcontents

\clearpage

\listoftables

\clearpage

\addtocontents{toc}{\protect\setcounter{tocdepth}{3}}
\addtocontents{lot}{\protect\setcounter{lotdepth}{3}}

\section{Dimensions of the Band Representations induced from the Wyckoff positions in paramagnetic space groups}\label{sec1}

For each (double) space group, the dimension of the band representation (BR) of $\mathcal{G}$ induced from the irrep $\rho_{\alpha}^i$ of the site symmetry group $\mathcal{G}_{\alpha}$ for each Wyckoff position (WP) $\alpha$ is given by

\begin{equation}
d(\rho_{\alpha}^i\uparrow\mathcal{G})=n_{\alpha}d(\rho_{\alpha}^i),
\label{Aeq1}
\end{equation} 
where $n_{\alpha}$ is the multiplicity of Wyckoff position $\alpha$ and $d(\rho_{\alpha}^i)$ is the dimension of the irrep $\rho_{\alpha}^i$.
Then, the  necessary and sufficient condition for a  TQC trivial insulator to be a filling enforced obstructed  atomic insulator (feOAI)  is
 \begin{equation}
\nexists N_{i,j} \in \mathbf{N}, s.t. N_e=\sum_{j=1}^{M}\sum_{i=1}^{N_{rep,\alpha_j}}d(\rho_{\alpha_j}^i\uparrow\mathcal{G})N_{i,j},
\label{Aeq2}  
\end{equation} 
where $N_e$ is the number of valence electrons and $\{\alpha_j\} (j=1,...,M)$ is the set of Wyckoff positions occupied by atoms. This condition guarantees that some of the the Wannier charge centers are localized at empty sites.

Using the theory of TQC and the \href{http://www.cryst.ehu.es}{Bilbao Crystallographic Server}, we have obtained the list of dimensions of all the induced BRs for each space group. See  subsections \ref{230SG} and \ref{230DSG} where we give the list of dimensions $d(\rho_{\alpha_j}^i\uparrow\mathcal{G})$ at each WP for all the SGs and DSGs; WPs sharing the same set of dimensions are grouped together.

\subsection{Properties of the dimensions of the BRs in the 230 paramagnetic SGs and DSGs}\label{app:properties}

Comparing the dimensions of the BRs induced from the WPs in each of the 230 SGs and DSGs yields the following general properties:
\begin{enumerate}
	\item At every WP of each SG and DSG, the dimensions of the BRs are multiples of the dimension of the smallest BR. Then for each WP we can just consider the BR with the smallest dimension at each WP in the tables of sections \ref{230SG} and \ref{230DSG}. This allows the grouping of different WPs, assigning the dimension of the smallest BR (DSBR) to each group of WPs.
	\item For all WPs in all SGs and DSGs, the dimension of the smallest induced BR with SOC is always twice the dimension of the smallest induced  BR without SOC. Therefore the conditions for the existence of a feOAI are identical for a SG and the corresponding DSG. 
	\item The smallest BR is always located at the $a$ position of each SG. For some SGs, there are also BRs with the same minimal dimension at other WPs.
	\item Except for the SGs 199, 214, 220 and 230, the dimension of the smallest BR at a given WP is always a multiple of the dimension of the smallest BR at WP $a$.
\end{enumerate}
As a consequence of these properties we can restrict the analysis of the feOAI conditions to the spinless case, where $N_e$ is one half the number of electrons.

\subsection{Dimensions of the BRs induced from the single SGs}\label{230SG}

Using the theory of TQC and the \href{http://www.cryst.ehu.es/cryst/bandrep}{BANDREP} tools at the \href{http://www.cryst.ehu.es}{BCS}, we have obtained the dimensions of all possible BRs induced from all the WPs in each SG without SOC  (see Table~\ref{tb:dimNoSOC}). In the following tables, different WPs that have the same multiplicity and the same DSBR are grouped together. 

\LTcapwidth=1.0\textwidth
\renewcommand\arraystretch{1.0}
\begin{scriptsize}

\begin{longtable}{|ll|l|l|l||ll|l|l|l|}
\caption[List of Wyckoff positions and band representation dimensions in each (single) space group]{List of Wyckoff positions (second column) and band representation dimensions (fourth column) in each (single) space group (first column). Maximal Wyckoff positions are typed in bold. The third and fourth columns give the multiplicity of the Wyckoff position in a primitive unit cell and the dimensions of the BRs induced from the Wyckoff position without SOC. Wyckoff position that share the same values of these two parameters are grouped together. \label{tb:dimNoSOC} } \\
\hline
\multicolumn{2}{|c|}{Space group}&Wyckoff pos.&m&d&\multicolumn{2}{|c|}{Space group}&Wyckoff pos.&m&d\\
\hline
\endfirsthead
\hline
\multicolumn{10}{|c|}{List of Wyckoff positions and band representation dimensions in each (single) space group}\\
\hline
\multicolumn{2}{|c|}{Space group}&Wyckoff pos.&m&d&\multicolumn{2}{|c|}{Space group}&Wyckoff pos.&m&d\\
\hline
\endhead
\hline
\rule{0pt}{10pt}
1&$P1$&$\mathbf{a}$&1&1&2&$P\bar{1}$&$\mathbf{a},\mathbf{b},\mathbf{c},\mathbf{d},\mathbf{e},\mathbf{f},\mathbf{g},\mathbf{h}$&1&1\\
&&&&&&&$i$&2&2\\
\hline\rule{0pt}{10pt}
3&$P2$&$\mathbf{a},\mathbf{b},\mathbf{c},\mathbf{d}$&1&1&4&$P2_1$&$\mathbf{a}$&2&2\\
&&$e$&2&2&&&&&\\
\hline\rule{0pt}{10pt}
5&$C2$&$\mathbf{a},\mathbf{b}$&1&1&6&$Pm$&$\mathbf{a},\mathbf{b}$&1&1\\
&&$c$&2&2&&&$c$&2&2\\
\hline\rule{0pt}{10pt}
7&$Pc$&$\mathbf{a}$&2&2&8&$Cm$&$\mathbf{a}$&1&1\\
&&&&&&&$b$&2&2\\
\hline
\end{longtable}
\end{scriptsize}

\subsection{Dimensions of the BRs induced from double space groups}\label{230DSG}

From the theory of TQC and the \href{http://www.cryst.ehu.es/cryst/bandrep}{BANDREP} tools at the \href{http://www.cryst.ehu.es}{BCS}, we have obtained the dimensions of all possible BRs induced from all the WPs in each DSG with SOC (see Table~\ref{tb:dimSOC}).  In the following tables, different WPs that have the same multiplicity and the same dimension of the smallest BR are grouped together.

\begin{scriptsize}

\begin{longtable}{|ll|l|l|l||ll|l|l|l|}
\caption[List of Wyckoff positions and band representation dimensions for each (double) space group]{List of Wyckoff positions (second column) and band representation dimensions (fourth column) for each (double) space group (first column). Maximal Wyckoff positions are typed in bold. The third and fourth columns give the multiplicity of the Wyckoff position in a primitive unit cell and the dimensions of the BRs induced from the Wyckoff position with SOC. Wyckoff position that share the same values of these two parameters are grouped together. } \\
\hline
\multicolumn{2}{|c|}{Space group}&Wyckoff pos.&m&d&\multicolumn{2}{|c|}{Space group}&Wyckoff pos.&m&d\\
\hline
\endfirsthead
\hline
\multicolumn{10}{|c|}{List of Wyckoff positions and band representation dimensions for each (double) space group}\\
\hline
\multicolumn{2}{|c|}{Space group}&Wyckoff pos.&m&d&\multicolumn{2}{|c|}{Space group}&Wyckoff pos.&m&d\\
\hline
\endhead
\rule{0pt}{10pt}
1&$P1$&$\mathbf{a}$&1&2&2&$P\bar{1}$&$\mathbf{a},\mathbf{b},\mathbf{c},\mathbf{d},\mathbf{e},\mathbf{f},\mathbf{g},\mathbf{h}$&1&2\\
&&&&&&&$i$&2&4\\
\hline\rule{0pt}{10pt}
3&$P2$&$\mathbf{a},\mathbf{b},\mathbf{c},\mathbf{d}$&1&2&4&$P2_1$&$\mathbf{a}$&2&4\\
&&$e$&2&4&&&&&\\
\hline\rule{0pt}{10pt}
5&$C2$&$\mathbf{a},\mathbf{b}$&1&2&6&$Pm$&$\mathbf{a},\mathbf{b}$&1&2\\
&&$c$&2&4&&&$c$&2&4\\
\hline\rule{0pt}{10pt}
7&$Pc$&$\mathbf{a}$&2&4&8&$Cm$&$\mathbf{a}$&1&2\\
&&&&&&&$b$&2&4\\
\end{longtable}
\end{scriptsize}

\section{Filling enforced conditions for the 230 DSGs}\label{fec_DSG}

Using the formula in Eq.~(\ref{Aeq2}) and the dimension of the \emph{smallest} BR (DSBR) of each group of WPs as tabulated in Section~\ref{230DSG}, we have obtained all the possible filling enforced conditions for each DSG depending on the occupied Wyckoff positions.  As tabulated in Table~\ref{fec_230DSG}, there are 13 SGs (1, 4, 7, 9, 19, 29, 33, 76, 78, 144, 145, 169 and 170) that have a unique DSBR implying the DSBR of the occupied sites and unoccupied sites are the same. Thus, the number of valence electrons of a topologically trivial insulator must be a multiple of the DSBR of the occupied sites and there are no filling enforced conditions for those groups.

On the other hand, 154 SGs fulfill the following condition: if we order the sets of WPs according to its DSBR (from lowest to highest), the DSBR of the $(i+1)^{\textrm{th}}$ set of WPs is a multiple of the DSBR of the $i^{\textrm{th}}$ set of WPs, i.e., DSBR$_{i+1}$/DSBR$_{i}\in\mathbb{N}$. In this case, the filling enforced condition is very easy to be derived: if the WP $a$ (with the minimal DSBR) is occupied by an atom, the material cannot be feOAI; otherwise, the minimal DSBR of the occupied WPs fixes the condition: if the number of electrons of a TQC trivial insulator, $N_e\ne \textrm{DSBR}_{\textrm{min}}\times\mathbb{N}$, then it is a Filling Enforced OAI.

For most SGs, the sufficient conditions in Eqs.~(\ref{eq5}) and (\ref{eq6}) are also the necessary conditions for a topologically trivial insulator to be a feOAI except for 29 SGs (162, 175, 189, 191, 195, 197, 200-202, 204, 207-209, 211, 214, 215, 217-219, 221-226, 228-230).

\LTcapwidth=1.0\textwidth
\renewcommand\arraystretch{1.0}
\begin{longtable}{|l|l|l|c|c|}
\caption[Filling enforced conditions for the 230 DSGs]{Filling enforced conditions for the 230 DSGs. For each SG we tabulate the set of WPs that must be empty in the 2nd column. In the 3rd column at least one of the WPs must be occupied. If the WPs in the 3rd column are grouped in brackets, \emph{at least} one of the WPs in each bracket must be occupied. The 4th column tabulates the necessary condition for a TQC trivial insulator. The 5th column is the corresponding filling condition for a TQC trivial insulator to be obstructed.} \\
\hline
SG&Empty&At least one Wyckoff&condition&condition\\
&&position occupied&TQC trivial&FEOAI\\
&&&insulator&\\
\hline
\endfirsthead
\hline
\multicolumn{5}{|c|}{Filling enforced conditions for the 230 DSGs}\\
\hline
SG&Empty&At least one Wyckoff&condition&condition\\
&&position occupied&TQC trivial&FEOAI\\
&&&insulator&\\
\hline
\endhead
\multirow{1}{*}{  1    $P1$}&\multicolumn{4}{c|}{No Filling Enforced OAI}\\
\hline
\multirow{1}{*}{  2    $P\bar{1}$}&$a,b,c,d,e,f,g,h$&$i$&$N_e=2\mathbb{N}$&$N_e=4\mathbb{N}+2$\\\cline{2-5}
\hline\multirow{1}{*}{  3    $P2$}&$a,b,c,d$&$e$&$N_e=2\mathbb{N}$&$N_e=4\mathbb{N}+2$\\\cline{2-5}
\hline\multirow{1}{*}{  4    $P2_1$}&\multicolumn{4}{c|}{No Filling Enforced OAI}\\
\hline
\multirow{1}{*}{  5    $C2$}&$a,b$&$c$&$N_e=2\mathbb{N}$&$N_e=4\mathbb{N}+2$\\\cline{2-5}
\hline\multirow{1}{*}{  6    $Pm$}&$a,b$&$c$&$N_e=2\mathbb{N}$&$N_e=4\mathbb{N}+2$\\\cline{2-5}
\hline\multirow{1}{*}{  7    $Pc$}&\multicolumn{4}{c|}{No Filling Enforced OAI}\\
\hline
\multirow{1}{*}{  8    $Cm$}&$a$&$b$&$N_e=2\mathbb{N}$&$N_e=4\mathbb{N}+2$\\
\end{longtable}

\section{FeOAIs obtained from High-throughput searches}\label{feoai_list}

In high-throughput search for  feOAIs we first select all the 34013 TQC topologically trivial insulators from the \webTQC. Then, for each material we obtain the number of electrons per unit cell and identify all the Wyckoff positions that are occupied by atoms.
Finally, by checking the filling enforced conditions in Eq.~\ref{Aeq2} and Table~\ref{fec_230DSG}, we determine if the material is a feOAI. We have found that 957 out of the 34013 TQC  trivial insulators satisfy the corresponding filling enforced conditions.
All the feOAIs are tabulated in Table~\ref{tab:oai} with their crystal and electronic properties, including the band gap, the number of valence electrons, the Wyckoff positions that are occupied and their corresponding DSBR.

\LTcapwidth=1.0\textwidth
\renewcommand\arraystretch{1.0}
\begin{longtable*}{|c|c|c|c|c|c|c|c|}
\caption[TQC topologically trivial insulators with filling enforced obstructed atomic insulating phase]{TQC topologically trivial insulators with filling enforced obstructed atomic insulating phase. For each material the table lists the space group symbol and number (SG), chemical formula, indirect band gap (IGap), direct band gap (DGap), number of valence electrons ($N_{e}$), occupied Wyckoff positions (WP$_{occ}$), the DSBR of each occupied Wyckoff position and all the related ICSD entries. Since the indirect band gap might slightly varies from one ICSD entry to another, IGap is the minimal indirect band gap for all entries and we provide in parentheses the indirect band gap for each ISCD.}\label{tab:oai} \\
\hline
No. & Formula & SG. & IGap(eV) & DGap(eV) & $N_{e}$ & {WP$_{occ}$(DSBR)} & ICSDs \\
\hline
\endfirsthead
\hline
\multicolumn{8}{|c|}{TQC topologically trivial insulators with filling enforced obstructed atomic insulating phase}\\
\hline
No. & Formula & SG. & IGap(eV) & DGap(eV) & $N_{e}$ & {WP$_{occ}$(DSBR)} & ICSDs \\
\hline
\endhead
{\scriptsize{1}} & {\scriptsize{ $\rm{Te}_{4} (\rm{Al}_{2} \rm{Cl}_{7})_{2}$}} & {\scriptsize{\sgsymb{2} (2)}} & {\scriptsize{0.032}} & {\scriptsize{0.032}} & {\scriptsize{134}} & {\scriptsize{i(4)}} & {\tiny{\icsdwebshort{426520}}} \\ \hline 
{\scriptsize{2}} & {\scriptsize{ $\rm{K} \rm{Fe} (\rm{P} \rm{O}_{4}) (\rm{N} \rm{O}_{3})$}} & {\scriptsize{\sgsymb{2} (2)}} & {\scriptsize{0.000}} & {\scriptsize{0.023}} & {\scriptsize{138}} & {\scriptsize{i(4)}} & {\tiny{\icsdwebshort{252847}}} \\ \hline 
{\scriptsize{3}} & {\scriptsize{ $\rm{Sr}_{10} (\rm{Mo}_{2} \rm{N}_{6}) (\rm{Mo} \rm{N}_{4})_{2}$}} & {\scriptsize{\sgsymb{2} (2)}} & {\scriptsize{1.236}} & {\scriptsize{1.236}} & {\scriptsize{194}} & {\scriptsize{i(4)}} & {\tiny{\icsdwebshort{413932}}} \\ \hline 
{\scriptsize{4}} & {\scriptsize{ $\rm{Cs}_{4} (\rm{P}_{2} \rm{Se}_{10})$}} & {\scriptsize{\sgsymb{2} (2)}} & {\scriptsize{1.524}} & {\scriptsize{1.613}} & {\scriptsize{106}} & {\scriptsize{i(4)}} & {\tiny{\icsdwebshort{418434}}} \\ \hline 
{\scriptsize{5}} & {\scriptsize{ $\rm{Te}_{4} (\rm{Sb} \rm{F}_{6})_{2}$}} & {\scriptsize{\sgsymb{2} (2)}} & {\scriptsize{1.064}} & {\scriptsize{1.286}} & {\scriptsize{118}} & {\scriptsize{i(4)}} & {\tiny{\icsdwebshort{201222}}} \\ \hline 
{\scriptsize{6}} & {\scriptsize{ $\rm{Hg}_{2} (\rm{Ga} \rm{Cl}_{4})_{2}$}} & {\scriptsize{\sgsymb{2} (2)}} & {\scriptsize{3.055}} & {\scriptsize{3.183}} & {\scriptsize{86}} & {\scriptsize{i(4)}} & {\tiny{\icsdwebshort{413579}}} \\ \hline 
{\scriptsize{7}} & {\scriptsize{ $\rm{Na}_{4} \rm{Sn}_{4} (\rm{C}_{2} \rm{O}_{4})_{3} \rm{F}_{6}$}} & {\scriptsize{\sgsymb{2} (2)}} & {\scriptsize{2.367}} & {\scriptsize{2.370}} & {\scriptsize{158}} & {\scriptsize{i(4)}} & {\tiny{\icsdwebshort{109824}}} \\ \hline 
{\scriptsize{8}} & {\scriptsize{ $\rm{Nb}_{2} \rm{O}_{2} \rm{F}_{3}$}} & {\scriptsize{\sgsymb{2} (2)}} & {\scriptsize{0.000}} & {\scriptsize{0.230}} & {\scriptsize{118}} & {\scriptsize{i(4)}} & {\tiny{\icsdwebshort{251695}}} \\
\hline
\end{longtable*}

\section{Surface states of feOAIs}\label{ss_calculations}

\subsection{Surface states of GaSe}

\begin{figure}[htbp]
\centering\includegraphics[width=5.2in]{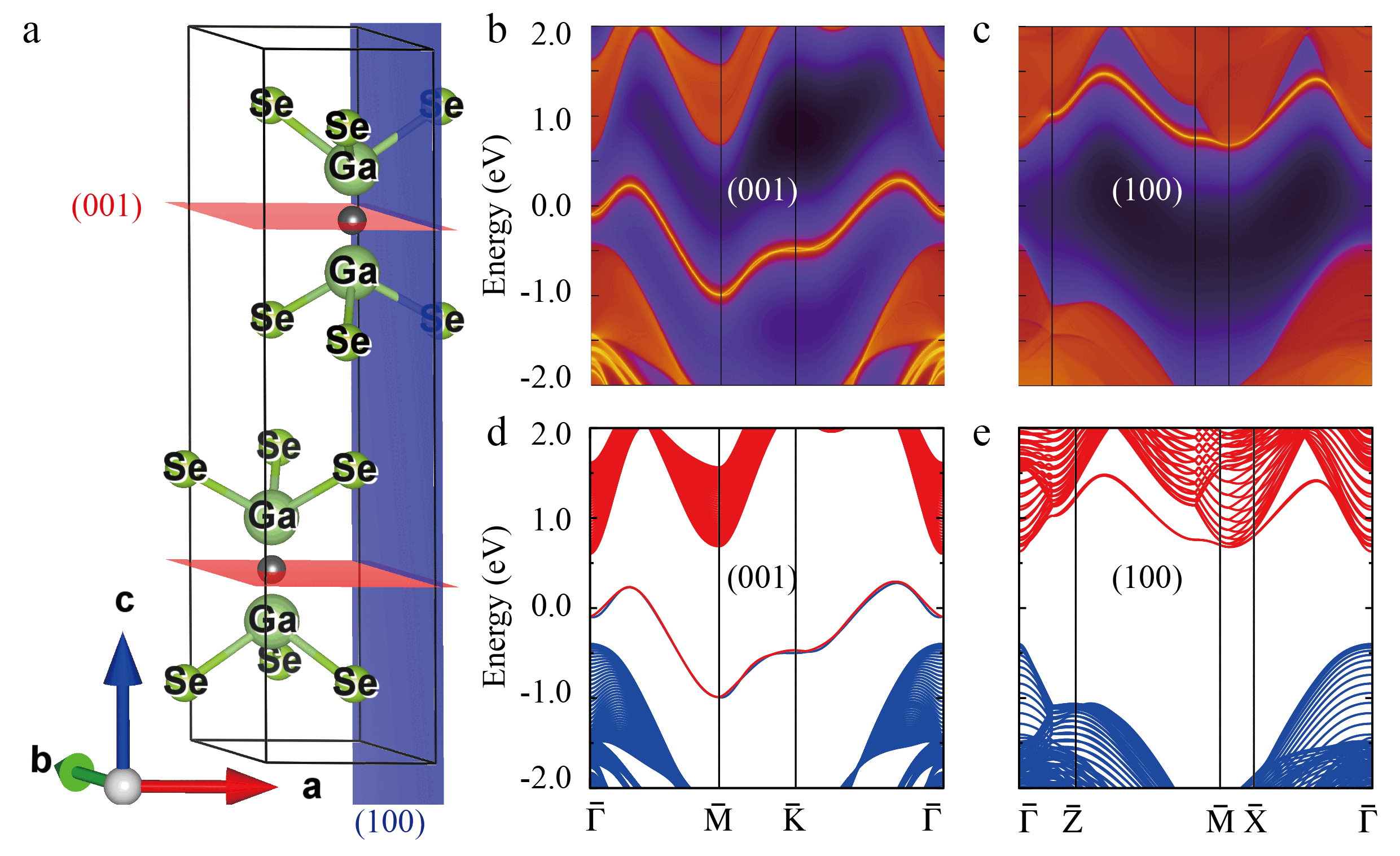}
\caption{(a) Crystal structure of GaSe, where the grey spheres show the positions of obstructed Wannier charge centers (OWCCs). 
The red and blue planes are the cleavage planes with Miller indices (100) and (001) respectively.
(b)-(c) Surface states for the semi-infinite structure with the cleavage planes defined in (a). The surface states are highlighted by the bright orange lines. (d)-(e) Surface states for the slab structure with the cleavage planes defined in (a).  In (d) and (e), by counting the number of bands from the lowest bands until charge neutrality, we indicate the valence and conduction bands by blue and red lines respectively. In (d) the highest valence band and the lowest conduction band are degenerate at the high-symmetry points, and the states on the (001) surface in (d) are metallic. On the other hand, the valence and conduction bands for the (100) surface are separated by a large band gap, and the (100) surface is insulating.
}\label{fig_GaSe}
\end{figure}

As shown in Figure~\ref{fig_GaSe}(a), GaSe with \icsdweb{20237} adopts a hexagonal lattice with space group  $P6_3/mmc$ (\#194). Both Ga and Se occupy the Wyckoff position $4f$. From the band structure in \webTQC (\icsdweb{20237}), GaSe is a TQC trivial insulator  with a direct gap of $1.175eV$.

The outer-shell electron configurations of Ga and Se are $4s^{2}4p^1$ and $4s^24p^4$ respectively.  As both Ga and Se sit at a Wyckoff position of multiplicity $4$, the total number of spinful electrons in one unit cell is $N_e=36$, which is divisible by $4$ but not by $8$. By checking the filling enforced conditions in Table~\ref{fec_230DSG} of Section~\ref{fec_DSG}, GaSe satisfies the filling enforced conditions of SG\#194, i.e. $N_e \in 8N+4$. Thus GaSe is a filling enforced obstructed atomic insulator.
 
From \webTQC (\icsdweb{20237}), the BR of GaSe (\ie{the complete set of bands below the Fermi level}) can be decomposed into linear combination of EBRs, then it was labeled as LCEBR in the \webTQC. In general, the decomposition is not unique. We find that the BR of GaSe can be decomposed in 24 ways, which are tabulated in Table~\ref{ebr_GaSe}. For each decomposition, the BR of irreps at $a, b, d$ and $g$ can always be moved to other Wyckoff positions. For instance, looking at the decompositions \#1 and \#2 in the table, the two orbitals at Wyckoff position $2d$ in \#2 are moved to Wyckoff position $2c$ in \#1 through the line $4f$ that connects $2c$ and $2d$, $\overline{E}_1@d\oplus\overline{E}_2@d\rightarrow\overline{E}_1@f\rightarrow\overline{E}_1@c\oplus\overline{E}_2@c$. However, in all the decompositions, there is always at least an orbital at Wyckoff position $2c$ (in fact the difference $m(\overline{E}_1@c)-m(\overline{E}_2@c)$ is 1 in all the decompositions). Therefore, an orbital $\overline{E}_1@c$ cannot be moved to other Wyckoff position without breaking the site symmetry of $2c$. This indicates that the BR of GaSe is necessarily obstructed at the unoccupied Wyckoff position $2c$, which results in metallic surface states on the cleavage plane that contains $2c$.
In Figure~\ref{fig_GaSe}(b) and (c) we show the surface states for GaSe on the (100) and (001) surfaces with a semi-infinite crystal structure computed by the Green's function method. As shown in Figure~\ref{fig_GaSe}(b) the surface states along the (001) direction are in the energy gap of bulk states. However the surface states on the (100) surface are above the chemical potential. 
To clarify the Fermi level of the surface states, we have also performed slab calculations. 
As shown in Figure~\ref{fig_GaSe}(d) and (e), the surface states from slab calculations are consistent with the surface states of the semi-infinite structure.
By counting the number of valence bands, which is equal to the number of electrons in one unit cell of the slab, from the lowest energy to charge neutrality, we find that the branch of connected surface states on the  (001) plane are half filled and hence metallic, which is also referred to as the \emph{filling anomaly}. In contrast, if the cleavage plane doesn't contain the OWCC, \ie $2c$ position, there is no filling anomaly on the surface.
For example, the (001) surface as indicated by the blue plane in Figure~\ref{fig_GaSe}(a) doesn't contain either the atoms or the $2c$ positions. In the surface states calculation, as plotted in Figure~\ref{fig_GaSe}(c) and (e), the valence bands and conduction bands are gapped and there is no filling anomaly.

\begin{sidewaystable}[h]
\caption[All possible decompositions of the BR of GaSe into linear combinations of EBRs for DSG \#194]{All possible decompositions of the BR of GaSe into linear combinations of EBRs for DSG \#194. The first row lists the EBRs induced from different orbitals at Wyckoff positions $\{a,b,c,d,g\}$; the numbers below are the coefficients of the EBRs for each particular decomposition. In each decomposition, the coefficient of $\overline{E}_1@c$ is always non-zero, indicating the BR of GaSe is obstructed at the unoccupied Wyckoff position $c$.}\label{ebr_GaSe}
\centering 
\begin{tabular}{c|c|c|c|c|c|c|c|c|c|c|c|c|c|c}
\hline\hline
case & $\overline{E}_1@b$ &$\overline{E}_2@b$ &$\overline{E}_{1u}@a$ &$\overline{E}_{1g}@a$ &$\overline{E}_{3}@d$/$\overline{E}_3@c$ &$\overline{E}_1@d$ &$\overline{E}_2@d$ &$\overline{E}_1@c$ &$\overline{E}_2@c$ &$\,^1\overline{E}_{u}\,^2\overline{E}_u@a$ &$\,^1\overline{E}_{g}\,^2\overline{E}_g@a$ &$\,^1\overline{E}_{u}\,^2\overline{E}_u@g$ &$\,^1\overline{E}_{g}\,^2\overline{E}_g@g$ &$\overline{E}_3@b$  \\
\hline
\#1 &0 &0 &0 &0 &0 &0 &0 &2 &1 &0 &0 &1 &1 &0  \\
\hline
\#2 &0 &0 &0 &0 &0 &1 &1 &1 &0 &0 &0 &1 &1 &0  \\
\hline
\#3 &0 &0 &0 &0 &2 &0 &0 &4 &3 &0 &0 &0 &0 &0  \\
\hline
\#4 &0 &0 &0 &0 &2 &1 &1 &3 &2 &0 &0 &0 &0 &0  \\
\hline
\#5 &0 &0 &0 &0 &2 &2 &2 &2 &1 &0 &0 &0 &0 &0  \\
\hline
\#6 &0 &0 &0 &0 &2 &3 &3 &1 &0 &0 &0 &0 &0 &0  \\
\hline
\#7 &0 &0 &1 &1 &1 &0 &0 &3 &2 &0 &0 &0 &0 &1  \\
\hline
\#8 &0 &0 &1 &1 &1 &1 &1 &2 &1 &0 &0 &0 &0 &1  \\
\hline
\#9 &0 &0 &1 &1 &1 &2 &2 &1 &0 &0 &0 &0 &0 &1  \\
\hline
\#10 &0 &0 &2 &2 &0 &0 &0 &2 &1 &0 &0 &0 &0 &2  \\
\hline
\#11 &0 &0 &2 &2 &0 &0 &0 &2 &1 &1 &1 &0 &0 &0  \\
\hline
\#12 &0 &0 &2 &2 &0 &1 &1 &1 &0 &0 &0 &0 &0 &2  \\
\hline
\#13 &0 &0 &2 &2 &0 &1 &1 &1 &0 &1 &1 &0 &0 &0  \\
\hline
\#14 &1 &1 &0 &0 &1 &0 &0 &3 &2 &0 &0 &0 &0 &1  \\
\hline
\#15 &1 &1 &0 &0 &1 &1 &1 &2 &1 &0 &0 &0 &0 &1  \\
\hline
\#16 &1 &1 &0 &0 &1 &2 &2 &1 &0 &0 &0 &0 &0 &1  \\
\hline
\#17 &1 &1 &1 &1 &0 &0 &0 &2 &1 &0 &0 &0 &0 &2  \\
\hline
\#18 &1 &1 &1 &1 &0 &0 &0 &2 &1 &1 &1 &0 &0 &0  \\
\hline
\#19 &1 &1 &1 &1 &0 &1 &1 &1 &0 &0 &0 &0 &0 &2  \\
\hline
\#20 &1 &1 &1 &1 &0 &1 &1 &1 &0 &1 &1 &0 &0 &0  \\
\hline
\#21 &2 &2 &0 &0 &0 &0 &0 &2 &1 &0 &0 &0 &0 &2  \\
\hline
\#22 &2 &2 &0 &0 &0 &0 &0 &2 &1 &1 &1 &0 &0 &0  \\
\hline
\#23 &2 &2 &0 &0 &0 &1 &1 &1 &0 &0 &0 &0 &0 &2  \\
\hline
\#24 &2 &2 &0 &0 &0 &1 &1 &1 &0 &1 &1 &0 &0 &0  \\
\hline\hline
\end{tabular}
\end{sidewaystable}

\subsection{Surface states of GaS} 

\begin{figure}[htbp]
\centering\includegraphics[width=5.2in]{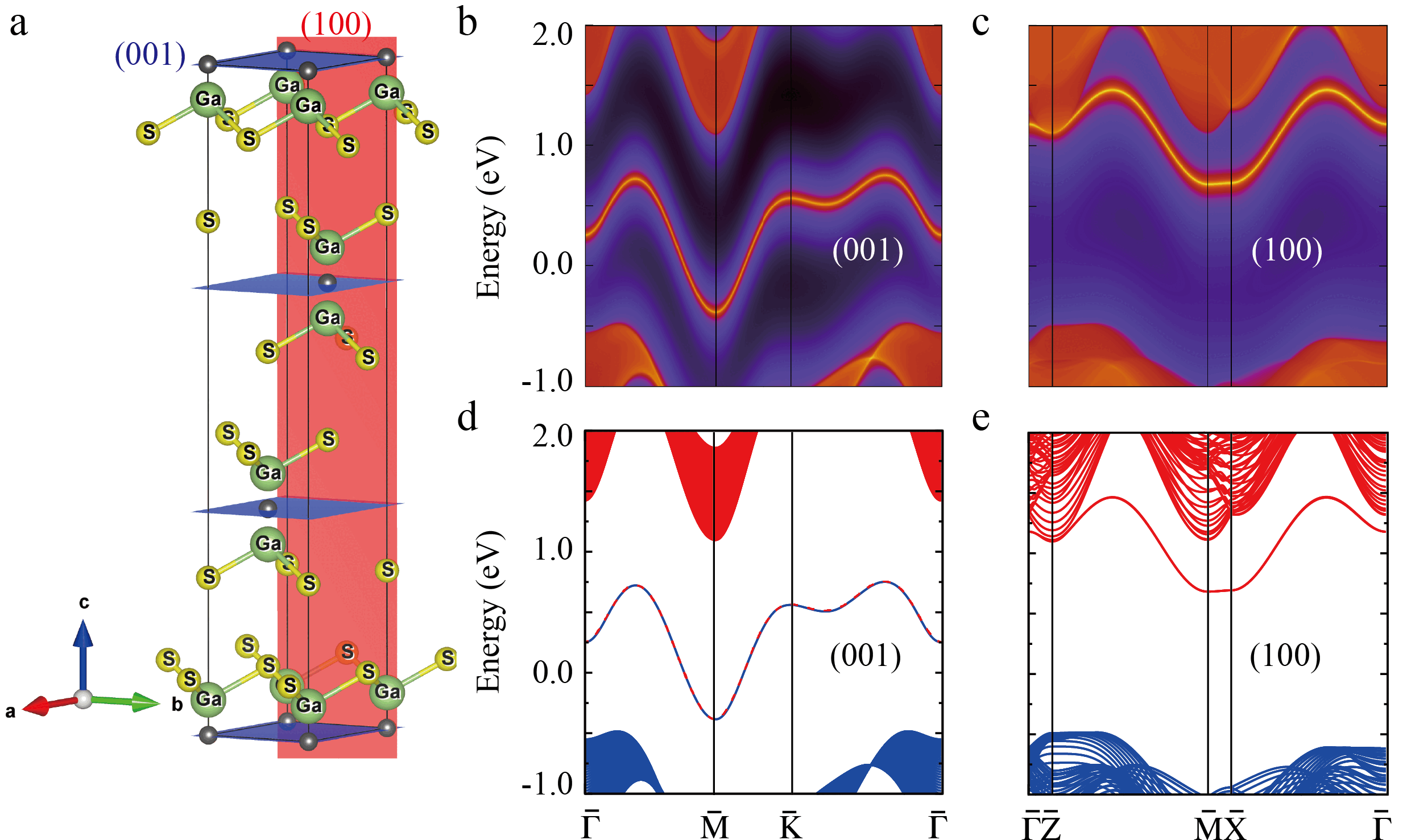}
\caption{(a) Crystal structure of GaS, where the grey spheres show  the positions of the OWCCs. 
The red and blue planes are the cleavage planes with Miller indices (100) and (001), respectively. 
(b)-(c) Surfaces states for a semi-infinite structure with the cleavage planes defined in (a). Surface states are highlighted by the bright orange lines. (d)-(e) Surfaces states for the the slab structure with the cleavage planes defined in (a).  In (d) and (e), by counting the number of bands from the lowest bands until charge neutrality, we indicate the valence and conduction bands by blue and red lines, respectively. In (d), the highest valence band and the lowest conduction band are degenerate at the high-symmetry points, and the (001) surface in (d) is metallic. On the other hand, the valence and conduction bands of (100) surface are separated by a large band gap, and the (100) surface is insulating.
}\label{fig_GaS}
\end{figure}

As shown in Figure~\ref{fig_GaS}(a), GaS with \icsdweb{40824} adopts a hexagonal lattice with trigonal space group  $R\bar3m$ (\#166) and it is a topologically trivial insulator with a direct gap of $1.5 eV$. The outer-shell electron configurations of Ga and S are $4s^{2}4p^1$ and $4s^24p^4$, respectively.  As both Ga and S sit at Wyckoff position $c$ with multiplicity $2$, the total number of spinful electrons in one unit cell is $N_e=18$, which satisfies the filling enforced conditions for SG\#166, i.e. $N_e \in 4N+2$. Thus, GaS is a filling enforced obstructed atomic insulator.
 
As in the analysis of GaSe, we find that the BR of GaS can be decomposed in 12 ways, which are tabulated in Table~\ref{ebr_GaS}. For each decomposition, the orbitals (irreps) at $b, d$ and $e$ can always be moved to some other Wyckoff positions. However, at least an orbital $\overline{E}_{1g}$ at the Wyckoff position $a$ is cannot be moved to another Wyckoff position.
Thus WP $1a$ is an OWCC. In Figure~\ref{fig_GaS}(b-e), we calculate the surface states of GaS on the (001) and (100) surfaces, as indicated by the blue and red planes in Figure~\ref{fig_GaS}(a) where the (001) plane contains an OWCC but the (100) plane does not, with both a semi-infinite and an infinite slab structures. 
By counting the number of valence bands, which is equal to the number of electrons in one unit cell of the slab, from the lowest energy to charge neutrality, we find that the branch of connected surface states on (001) plane are half filled and hence there is a filling anomaly. By contrast, the surface states on the (100) plane, which doesn't contain the OWCC at $a$, are insulating. 

\begin{sidewaystable}[h]
\caption[All possible decompositions of the BR of GaS into linear combinations of EBRs for DSG \#166]{All possible decompositions of the BR of GaS into linear combinations of EBRs for DSG \#166. The first row lists the EBRs induced from different orbitals at Wyckoff positions $\{a,b,d,g\}$; the numbers below are the coefficients of the EBRs for each particular decomposition. In each decomposition, the coefficient of $\bar{E_{1u}}@a$ is always non-zero, indicating the BR of GaSe is obstructed at the unoccupied Wyckoff position $a$.}\label{ebr_GaS}
\centering 
\begin{tabular}{c|c|c|c|c|c|c|c|c|c|c|c|c}
\hline\hline
case & $\overline{E}_{1u}@b$ &$\overline{E}_{1g}@b$ &$\overline{E}_{1u}@a$ &$\overline{E}_{1g}@a$ &$\,^1\overline{E}_{u}\,^2\overline{E}_{u}@b$&$\,^1\overline{E}_{g}\,^2\overline{E}_{g}@b$&$\,^1\overline{E}_{u}\,^2\overline{E}_{u}@a$ &$\,^1\overline{E}_{g}\,^2\overline{E}_{g}@a$&$\,^1\overline{E}_{u}\,^2\overline{E}_{u}@e$ &
$\,^1\overline{E}_{g}\,^2\overline{E}_{g}@e$ & $\,^1\overline{E}_{u}\,^2\overline{E}_{u}@d$ &$\,^1\overline{E}_{g}\,^2\overline{E}_{g}@d$  \\
\hline
 \#1 & 0 & 0 & 1 & 2 & 0 & 0 & 0 & 0 & 0 & 0 & 1 & 1 \\
\hline
 \#2 & 0 & 0 & 1 & 2 & 0 & 0 & 0 & 0 & 1 & 1 & 0 & 0 \\
\hline
 \#3 & 0 & 0 & 3 & 4 & 0 & 0 & 1 & 1 & 0 & 0 & 0 & 0 \\
\hline
 \#4 & 0 & 0 & 3 & 4 & 1 & 1 & 0 & 0 & 0 & 0 & 0 & 0 \\
\hline
 \#5 & 1 & 1 & 0 & 1 & 0 & 0 & 0 & 0 & 0 & 0 & 1 & 1 \\
\hline
 \#6 & 1 & 1 & 0 & 1 & 0 & 0 & 0 & 0 & 1 & 1 & 0 & 0 \\
\hline
 \#7 & 1 & 1 & 2 & 3 & 0 & 0 & 1 & 1 & 0 & 0 & 0 & 0 \\
\hline
 \#8 & 1 & 1 & 2 & 3 & 1 & 1 & 0 & 0 & 0 & 0 & 0 & 0 \\
\hline
 \#9 & 2 & 2 & 1 & 2 & 0 & 0 & 1 & 1 & 0 & 0 & 0 & 0 \\
\hline
 \#10& 2 & 2 & 1 & 2 & 1 & 1 & 0 & 0 & 0 & 0 & 0 & 0 \\
\hline
 \#11& 3 & 3 & 0 & 1 & 0 & 0 & 1 & 1 & 0 & 0 & 0 & 0 \\
\hline
 \#12& 3 & 3 & 0 & 1 & 1 & 1 & 0 & 0 & 0 & 0 & 0 & 0 \\
\hline\hline
\end{tabular}
\label{tab:LPer}
\end{sidewaystable}

\color{black}

\subsection{Surface states of IrP$_2$}

\begin{figure}[htbp]
\centering\includegraphics[width=5.2in]{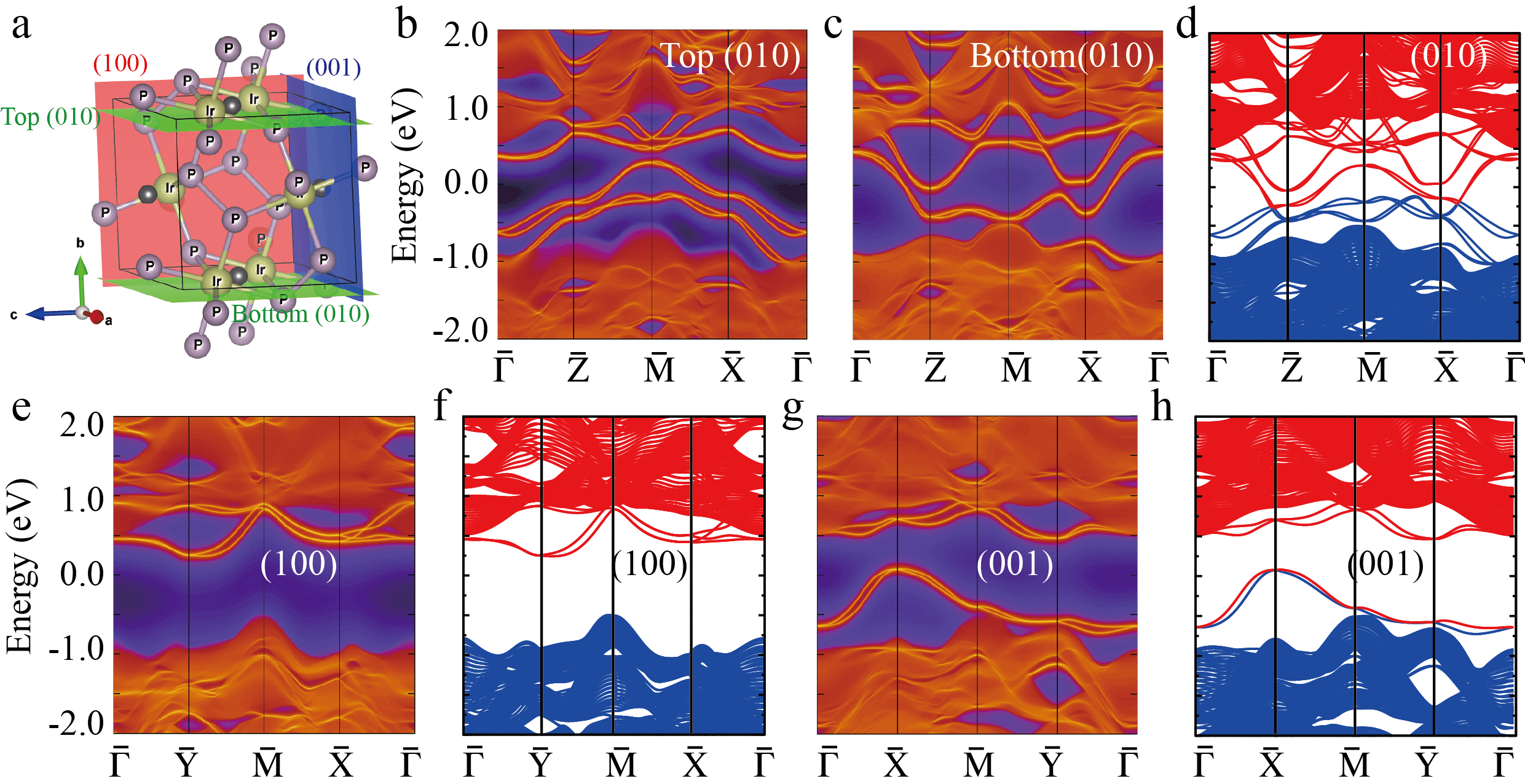}
\caption{(a) Crystal structure of IrP$_2$, where the black spheres show  the positions of OWCCs. 
The red, green and blue planes are the cleavage planes with Miller indices (100), (010) and (001), respectively.
(b)-(d) Surface states  on the (010) plane with the semi-infinite structure and slab structure. The surface states on the top (010) surface in (b) and bottom (010) surface in (c) have an accidental crossing in the slab calculation in (d).
(e)-(f) Surface states  on the (100) plane with the semi-infinite structure and slab structure.
(g)-(h) Surface states  on the (001) plane with the semi-infinite structure and slab structure.
By counting the number of bands from the lowest bands till charge neutrality, we indicate the valence and conduction bands by blue and red lines, respectively.
 In (f), the highest valence band and the lowest conduction band are separated by a large band gap, and the (100) surface is insulating. On the other hand, the valence and conduction bands on the (001) surface are degenerate at the high-symmetry points, and the (001) surface in (d) is metallic.}\label{fig_IrP2}
\end{figure}

IrP$_2$ with \icsdweb{44661} is a TQC trivial insulator with an indirect band gap of $1.0 eV$
As shown in Figure~\ref{fig_IrP2}(a), the crystal structure of IrP2 adopts the space group $P2_1/c$ (\#14) with both Ir and P occupying the WP $4e$. As the outer-shell electron configurations of Ir and P are $5d^{7}6s^2$ and $3s^23p^3$, respectively, the total number of electrons in one unit cell is $N_e=76$, which also satisfies the filling enforced conditions for SG\#14, i.e. $N_e \in 8N+4$. Thus IrP$_2$ is a filling enforced obstructed atomic insulator.
By analyzing the BR of IrP$_2$ and its decomposition into the EBRs of SG\#14 as in the previous examples, we find that the Wyckoff position $2d$
is necessarily an OWCC. 

In Figure~\ref{fig_IrP2}(a), the (100) and (010) cleavage planes are away from the OWCC at WP $2d$, while the (001) cleavage plane cuts the OWCC. It should be noted that the OWCC is located on the same plane with Miller indices (010) as the Ir atoms, and then it is not possible to have a surface which contains the OWCC and not the Ir atoms.  One possible cleavage plane on (010)
 surface is the green plane, which is next to the OWCC and indicated in Figure~\ref{fig_IrP2}(a). Since it doesn't contain any OWCC, there is no filling anomaly on this plane. As shown in Figure~\ref{fig_IrP2}(d), although several crossing points appear near the Fermi level of the slab band structure of this cleavage plane, they come from the accidental band crossing between the surface bands of top and bottom surfaces, which are also calculated in Figures~\ref{fig_IrP2} (b) and~(c).  As the top and bottom surfaces are not related by symmetry, the band crossing can be removed by surface potential giving an insulating surface. The surface states of another cleavage on the (100) plane, \ie the red plane in Figure~\ref{fig_IrP2}(a), are also calculated and shown in Figure~\ref{fig_IrP2}(e) and (f). As it doesn't contain any OWCC either, there is no filling anomaly in its slab band structure.
 
Different from the above two Miller planes, it is possible to have a cleavage that cuts the OWCC at $2d$ position but it is away from the atoms on the (001) plane. This surface is indicated by the blue plane in Figure~\ref{fig_IrP2}(a). Thus, this cleavage plane has filling anomaly. As shown in Figure~\ref{fig_IrP2}(g) and (h), by counting the number of valence bands, the surface states in the bulk band gap are half-filled and hence are metallic.

\subsection{Surface states of CuP$_2$}

\begin{figure}[htbp]
\centering\includegraphics[width=5.2in]{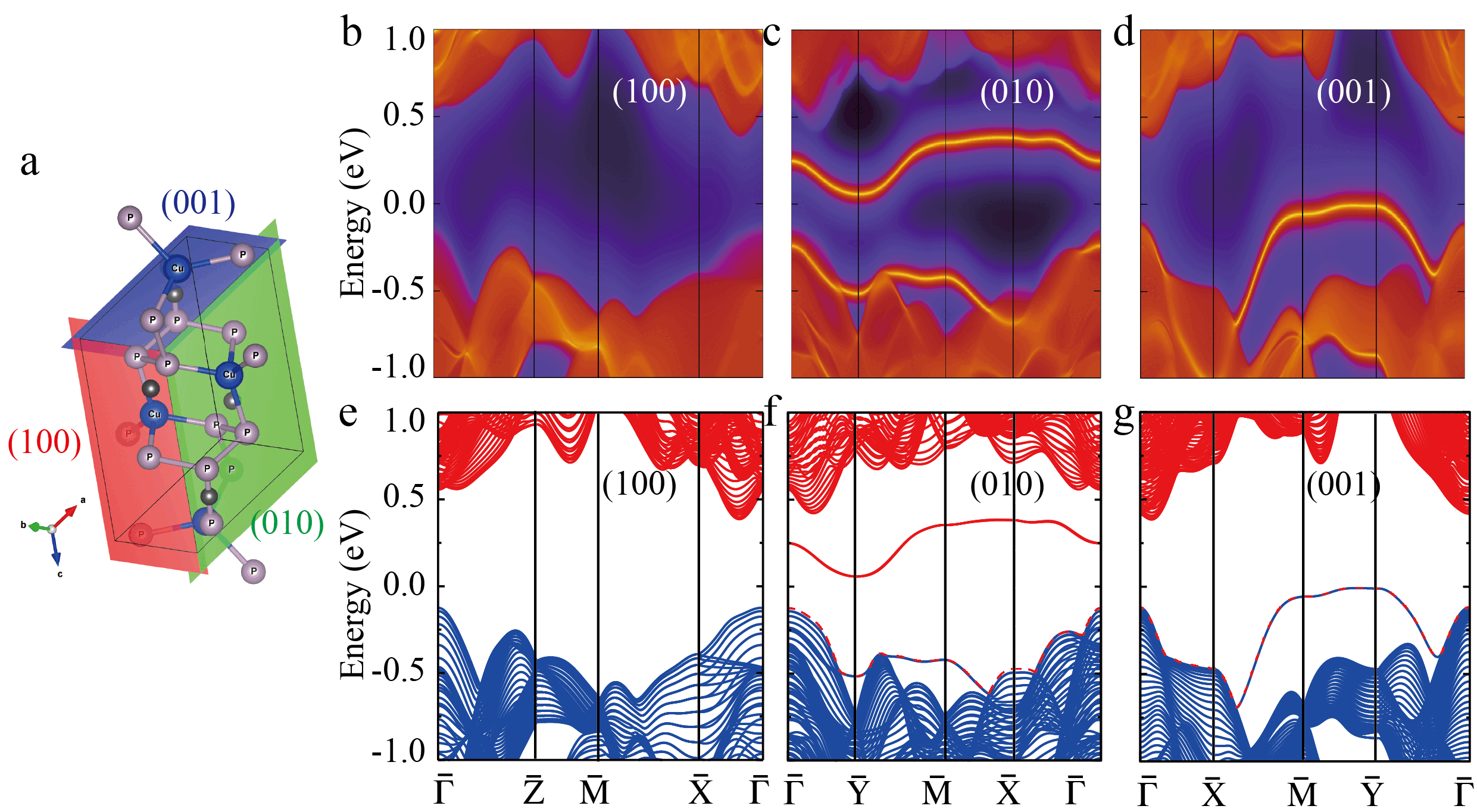}
\caption{(a) Crystal structure of CuP$_2$, where the black spheres indicate the positions of the  OWCCs 
The red, green and blue planes are the cleavage planes with Miller indices (100), (010) and (001), respectively  
(b)-(d) Surface states for the semi-infinite structure with the cleavage planes defined in (a). Surface states are highlighted by the bright orange lines. (e)-(g) Surface states for the slab structure with the cleavage planes defined in (a).  By counting the number of bands from the lowest bands until charge neutrality, we plot the valence and conduction bands by blue and red lines in the slab calculations, respectively. In (e), the valence and conduction bands of (100) surface are separated by a large band gap. Thus the (100) surface is insulating, 
while in (f) and (g) the highest valence band and the lowest conduction band are degenerate at the high-symmetry points, and the (010) and (001) surfaces are metallic.}\label{fig_CuP2}
\end{figure}

Like  IrP$_2$, CuP$_2$ with \icsdweb{35282} is a TQC topologically trivial insulator with an indirect band gap of $0.85 eV$, 
and its crystal structure adopts the space group $P2_1/c$ (\#14) with both Cu and P occupying the WP $4e$, as shown in Figure~\ref{fig_CuP2}(a). The outer-shell electron configuration of Cu and P are $3d^{10}4s^1$ and $3s^23p^3$, respectively. Thus there are 84 valence electrons in one unit cell, which satisfies the filling enforced conditions for SG\#14, i.e. $N_e \in 8N+4$. Hence, CuP$_2$ is a filling enforced obstructed atomic insulator.
The analysis of the decomposition of the BR of CuP$_2$ into  EBRs of SG\#14 shows that the Wyckoff position $2d$
is necessarily an OWCC.

In Figure~\ref{fig_CuP2}(a) the (100) cleavage planes are away from the OWCC at WP $2d$, while the (010) and (001) cleavage planes contain the OWCC. In Figure~\ref{fig_IrP2}(b-d) we calculate the surface states of CuP$_2$ on the (100), (010) and (001) planes. As the (010) and (001) cleavage planes contain the OWCC, counting the number of valence bands, which is equal to the number of electrons in one unit cell of the slab, shows that the (001) surface is metallic, see  Figure~\ref{fig_CuP2}(f) and (g). In contrast, the states on the (100) plane which does not contain any OWCC in Figures~\ref{fig_CuP2}(b) and~(e) are insulating. 

\section{Magnetic feOAIs: Er$_2$Ni$_2$In}\label{mag_feoais}

The magnetic crystal structure of Er$_2$Ni$_2$In adopts the Type-IV MSG $C_amcm$ (\#63.467) (\url{http://webbdcrista1.ehu.es/magndata/index.php?this_label=1.195}). In the first-principle calculations, when we take the Hubbard-U parameter of $3d$ electrons on Ni as $U=2 eV$ and $4f$ electrons on Er as $U=0,1,5,6 eV$, the occupied bands are classified as topologically trivial insulator. As tabulated in Table~\ref{mag_1.195}, the occupied Wyckoff positions are $\{e,i,j\}$. There are 356 valence electrons in one unit cell, which satisfies the the filling enforced condition of MSG $C_amcm$ (\#63.467), $i.e.$, $N_e=8N+4$, as tabulated in Table~\ref{tb:dimTypeIVSOC}.

\begin{table}
\centering
\begin{tabular}{c|c|c|c}
\hline\hline
Atom &  Number of outer-shell electrons & Wyckoff position  & Site Sym. \\
\hline
Er  & 22 &  $8j$  &  $m'$ \\
\hline
Ni  & 16  &  $8i$   &  $m$ \\
\hline
In  & 13  &  $4e$    &  $m2m$ \\
\hline\hline
\end{tabular}
\caption[The occupied Wyckoff positions in the magnetic material Er$_2$Ni$_2$In with MSG $C_amcm$ (\#63.467)]{The occupied Wyckoff positions in the magnetic material Er$_2$Ni$_2$In with MSG $C_amcm$ (\#63.467).}\label{mag_1.195}
\end{table}

\clearpage

\section{Filling enforced conditions for magnetic space groups}\label{fec_1651SSG}

In this section we extend the formula in Eq.~(\ref{Aeq2}) to the 1651 SSGs, which include 230 Type-I magnetic space groups (MSGs), 230 Type-II DSGs, 674 Type-III MSGs and 517 Type-IV MSGs. The 230 Type-II DSGs are paramagnetic groups, 
for which the filling enforced conditions have been obtained in Section~\ref{fec_DSG}. The remaining 1421 groups are MSGs. The magnetic Wyckoff positions and their dimensions have been studied recently in the context of magnetic topological quantum chemistry (MTQC) \cite{MTQC,xu2020high}. The Magnetic Band Representations (MBRs) induced from all the Wyckoff positions can be obtained in the \href{http://www.cryst.ehu.es/cryst/mbandrep}{MBANREP} tool at the \href{http://www.cryst.ehu.es/cryst/mbandrep}{BCS}. 
For completeness, in sections \ref{dimTypeINoSOC}-\ref{dimTypeIVSOC} we give the multiplicity of each Wyckoff position in a primitive cell and the dimension of the \emph{smallest} MBR (DSMBR) induced from that Wyckoff position. Afterwards, in sections \ref{fecTypeINoSOC}-\ref{fecTypeIVSOC} we tabulate the filling enforced conditions for the existence of magnetic obstructed atomic insulators in each MSG. For a MSG, if there is only a unique DSMBR, it implies that the DSMBR of the occupied sites and unoccupied sites are the same. Thus, the number of valence electrons of a topologically trivial magnetic insulator must be a multiple of the DSMBR of the occupied sites and there is no filling enforced conditions for this MSG.
In non-magnetic SGs, the DSBR with SOC at a given Wyckoff position is always 2 times the DSBR without SOC at the same Wyckoff position. Unlike non-magnetic SGs, although the DSMBR with SOC of a Wyckoff position in MSG is a multiple of the DSMBR without SOC at the same Wyckoff position, the multiple is not 2 in all the Wyckoff positions. For example, in MSG\#22.45($F222$), the DSMBR with SOC at $1a$ is two times of the DSMBR without SOC at $1a$. However the DSMBR both with and without SOC at $2e$ are identical.  Thus the filling enforced conditions are different for this MSG with and without SOC. In the following, we include in different tables the conditions for magnetic systems with and without SOC.

Note that in the absence of SOC, the single-valued irreps of MSGs can be used only when the Hamiltonian does not depend on the spin of the electrons. In magnetic materials without SOC, in general, only in antiferromagnetic systems can be assumed Hamiltonians that do not depend on the spin degrees of freedom but not in ferromagnetic systems.  Indeed, the Hamiltonian of ferromagnetic material with spin polarization does depend on the spin. So the filling enforced conditions of MSGs without SOC can only be applied to antiferromagnetic topologically trivial insulators. 

\subsection{Dimensions of the BRs for each WP in single (without SOC) SSGs of Type I}\label{dimTypeINoSOC}

\begin{longtable}{|ll|l|l|l||ll|l|l|l|}
\caption[List of Wyckoff positions and BR dimensions in Type-I SSGs without SOC]{List of Wyckoff positions (second column) and BR dimensions  (fourth column) in Type-I SSGs without SOC (first column). Maximal Wyckoff positions are typed in bold. The third and fourth columns give the multiplicity of the Wyckoff position in a primitive unit cell and the dimensions of the BRs induced from the Wyckoff position without SOC. Wyckoff position that share the same values of these two parameters are grouped together. } \\
\hline
\multicolumn{2}{|c|}{Magnetic group}&Wyckoff pos.&m&d&\multicolumn{2}{|c|}{Magnetic group}&Wyckoff pos.&m&d\\
\hline
\endfirsthead
\hline
\multicolumn{10}{|c|}{List of Wyckoff positions and BR dimensions in Type-I SSGs without SOC}\\
\hline
\multicolumn{2}{|c|}{Magnetic group}&Wyckoff pos.&m&d&\multicolumn{2}{|c|}{Magnetic group}&Wyckoff pos.&m&d\\
\hline
\endhead
\rule{0pt}{10pt}
1.1&$P1$&$\mathbf{a}$&1&1&2.4&$P\bar{1}$&$\mathbf{a},\mathbf{b},\mathbf{c},\mathbf{d},\mathbf{e},\mathbf{f},\mathbf{g},\mathbf{h}$&1&1\\
&&&&&&&$i$&2&2\\
\hline
\rule{0pt}{10pt}
3.1&$P2$&$\mathbf{a},\mathbf{b},\mathbf{c},\mathbf{d}$&1&1&4.7&$P2_1$&$\mathbf{a}$&2&2\\
&&$e$&2&2&&&&&\\
\hline
\rule{0pt}{10pt}
5.13&$C2$&$\mathbf{a},\mathbf{b}$&1&1&6.18&$Pm$&$\mathbf{a},\mathbf{b}$&1&1\\
&&$c$&2&2&&&$c$&2&2\\
\hline
\rule{0pt}{10pt}
7.24&$Pc$&$\mathbf{a}$&2&2&8.32&$Cm$&$\mathbf{a}$&1&1\\
&&&&&&&$b$&2&2\\
\end{longtable}

\subsection{Dimensions of the BRs for each WP in single SSGs (without SOC) of Type III}\label{dimTypeIIINoSOC}

\begin{longtable}{|ll|l|l|l||ll|l|l|l|}
\caption[List of Wyckoff positions and BR dimensions in Type-III SSGs without SOC]{List of Wyckoff positions (second column) in Type-III SSGs without SOC (first column). Maximal Wyckoff positions are typed in bold. The third and fourth columns give the multiplicity of the Wyckoff position in a primitive unit cell and the dimensions of the BRs induced from the Wyckoff position without SOC. Wyckoff position that share the same values of these two parameters are grouped together. } \\
\hline
\multicolumn{2}{|c|}{Magnetic group}&Wyckoff pos.&m&d&\multicolumn{2}{|c|}{Magnetic group}&Wyckoff pos.&m&d\\
\hline
\endfirsthead
\hline
\multicolumn{10}{|c|}{List of Wyckoff positions and BR dimensions in Type-III SSGs without SOC}\\
\hline
\multicolumn{2}{|c|}{Magnetic group}&Wyckoff pos.&m&d&\multicolumn{2}{|c|}{Magnetic group}&Wyckoff pos.&m&d\\
\hline
\endhead
\rule{0pt}{10pt}
2.6&$P\bar{1}'$&$\mathbf{a},\mathbf{b},\mathbf{c},\mathbf{d},\mathbf{e},\mathbf{f},\mathbf{g},\mathbf{h}$&1&1&3.3&$P2'$&$\mathbf{a},\mathbf{b},\mathbf{c},\mathbf{d}$&1&1\\
&&$i$&2&2&&&$e$&2&2\\
\hline
\rule{0pt}{10pt}
4.9&$P2_1'$&$\mathbf{a}$&2&2&5.15&$C2'$&$\mathbf{a},\mathbf{b}$&1&1\\
&&&&&&&$c$&2&2\\
\hline
\rule{0pt}{10pt}
6.20&$Pm'$&$\mathbf{a},\mathbf{b}$&1&1&7.26&$Pc'$&$\mathbf{a}$&2&2\\
&&$c$&2&2&&&&&\\
\hline
\rule{0pt}{10pt}
8.34&$Cm'$&$\mathbf{a}$&1&1&9.39&$Cc'$&$\mathbf{a}$&2&2\\
&&$b$&2&2&&&&&\\
\end{longtable}

\subsection{Dimensions of the BRs for each WP in single SSGs (without SOC) of Type IV}\label{dimTypeIVNoSOC}
\begin{longtable}{|ll|l|l|l||ll|l|l|l|}
\caption[List of Wyckoff positions and BR dimensions in Type-IV SSGs without SOC]{List of Wyckoff positions (second column) in Type-IV SSGs without SOC (first column). Maximal Wyckoff positions are typed in bold. The third and fourth columns give the multiplicity of the Wyckoff position in a primitive unit cell and the dimensions of the BRs induced from the Wyckoff position without SOC. Wyckoff position that share the same values of these two parameters are grouped together. } \\
\hline
\multicolumn{2}{|c|}{Magnetic group}&Wyckoff pos.&m&d&\multicolumn{2}{|c|}{Magnetic group}&Wyckoff pos.&m&d\\
\hline
\endfirsthead
\hline
\multicolumn{10}{|c|}{List of Wyckoff positions and BR dimensions in Type-IV SSGs without SOC}\\
\hline
\multicolumn{2}{|c|}{Magnetic group}&Wyckoff pos.&m&d&\multicolumn{2}{|c|}{Magnetic group}&Wyckoff pos.&m&d\\
\hline
\endhead
\rule{0pt}{10pt}
1.3&$P_S1$&$\mathbf{a}$&2&2&2.7&$P_S\bar{1}$&$\mathbf{a},\mathbf{b},\mathbf{c},\mathbf{d},\mathbf{e},\mathbf{f},\mathbf{g},\mathbf{h}$&2&2\\
&&&&&&&$i$&4&4\\
\hline
\rule{0pt}{10pt}
3.4&$P_a2$&$\mathbf{a},\mathbf{b},\mathbf{c},\mathbf{d}$&2&2&3.5&$P_b2$&$\mathbf{a},\mathbf{b},\mathbf{c},\mathbf{d}$&2&2\\
&&$e$&4&4&&&$e$&4&4\\
\hline
\rule{0pt}{10pt}
3.6&$P_C2$&$\mathbf{a},\mathbf{b}$&2&2&4.10&$P_a2_1$&$\mathbf{a}$&4&4\\
&&$c$&4&4&&&&&\\
\hline
\rule{0pt}{10pt}
4.11&$P_b2_1$&$\mathbf{a},\mathbf{b},\mathbf{c},\mathbf{d}$&2&2&4.12&$P_C2_1$&$\mathbf{a},\mathbf{b}$&2&2\\
&&$e$&4&4&&&$c$&4&4\\
\end{longtable}

\subsection{Dimensions of the BRs for each WP in double SSGs (with SOC) of Type I}\label{dimTypeISOC}
\begin{longtable}{|ll|l|l|l||ll|l|l|l|}
\caption[List of Wyckoff positions and BR dimensions in Type-I SSGs with SOC]{List of Wyckoff positions (second column) in Type-I SSGs with SOC (first column). Maximal Wyckoff positions are typed in bold. The third and fourth columns give the multiplicity of the Wyckoff position in a primitive unit cell and the dimensions of the BRs induced from the Wyckoff position with SOC. Wyckoff position that share the same values of these two parameters are grouped together. } \\
\hline
\multicolumn{2}{|c|}{Magnetic group}&Wyckoff pos.&m&d&\multicolumn{2}{|c|}{Magnetic group}&Wyckoff pos.&m&d\\
\hline
\endfirsthead
\hline
\multicolumn{10}{|c|}{List of Wyckoff positions and BR dimensions in Type-I SSGs with SOC}\\
\hline
\multicolumn{2}{|c|}{Magnetic group}&Wyckoff pos.&m&d&\multicolumn{2}{|c|}{Magnetic group}&Wyckoff pos.&m&d\\
\hline
\endhead
\rule{0pt}{10pt}
1.1&$P1$&$\mathbf{a}$&1&1&2.4&$P\bar{1}$&$\mathbf{a},\mathbf{b},\mathbf{c},\mathbf{d},\mathbf{e},\mathbf{f},\mathbf{g},\mathbf{h}$&1&1\\
&&&&&&&$i$&2&2\\
\hline
\rule{0pt}{10pt}
3.1&$P2$&$\mathbf{a},\mathbf{b},\mathbf{c},\mathbf{d}$&1&1&4.7&$P2_1$&$\mathbf{a}$&2&2\\
&&$e$&2&2&&&&&\\
\hline
\rule{0pt}{10pt}
5.13&$C2$&$\mathbf{a},\mathbf{b}$&1&1&6.18&$Pm$&$\mathbf{a},\mathbf{b}$&1&1\\
&&$c$&2&2&&&$c$&2&2\\
\hline
\rule{0pt}{10pt}
7.24&$Pc$&$\mathbf{a}$&2&2&8.32&$Cm$&$\mathbf{a}$&1&1\\
&&&&&&&$b$&2&2\\
\end{longtable}

\subsection{Dimensions of the BRs for each WP in double SSGs (with SOC) of Type III}\label{dimTypeIIISOC}
\begin{longtable}{|ll|l|l|l||ll|l|l|l|}
\caption[List of Wyckoff positions and BR dimensions in Type-III SSGs with SOC]{List of Wyckoff positions (second column) in Type-III SSGs with SOC (first column). Maximal Wyckoff positions are typed in bold. The third and fourth columns give the multiplicity of the Wyckoff position in a primitive unit cell and the dimensions of the BRs induced from the Wyckoff position with SOC. Wyckoff position that share the same values of these two parameters are grouped together. } \\
\hline
\multicolumn{2}{|c|}{Magnetic group}&Wyckoff pos.&m&d&\multicolumn{2}{|c|}{Magnetic group}&Wyckoff pos.&m&d\\
\hline
\endfirsthead
\hline
\multicolumn{10}{|c|}{List of Wyckoff positions and BR dimensions in Type-III SSGs with SOC}\\
\hline
\multicolumn{2}{|c|}{Magnetic group}&Wyckoff pos.&m&d&\multicolumn{2}{|c|}{Magnetic group}&Wyckoff pos.&m&d\\
\hline
\endhead
\rule{0pt}{10pt}
2.6&$P\bar{1}'$&$\mathbf{a},\mathbf{b},\mathbf{c},\mathbf{d},\mathbf{e},\mathbf{f},\mathbf{g},\mathbf{h}$&1&2&3.3&$P2'$&$\mathbf{a},\mathbf{b},\mathbf{c},\mathbf{d}$&1&1\\
&&$i$&2&2&&&$e$&2&2\\
\hline
\rule{0pt}{10pt}
4.9&$P2_1'$&$\mathbf{a}$&2&2&5.15&$C2'$&$\mathbf{a},\mathbf{b}$&1&1\\
&&&&&&&$c$&2&2\\
\hline
\rule{0pt}{10pt}
6.20&$Pm'$&$\mathbf{a},\mathbf{b}$&1&1&7.26&$Pc'$&$\mathbf{a}$&2&2\\
&&$c$&2&2&&&&&\\
\hline
\rule{0pt}{10pt}
8.34&$Cm'$&$\mathbf{a}$&1&1&9.39&$Cc'$&$\mathbf{a}$&2&2\\
&&$b$&2&2&&&&&\\
\end{longtable}

\subsection{Dimensions of the BRs for each WP in double SSGs (with SOC) of Type IV}\label{dimTypeIVSOC}
\begin{longtable}{|ll|l|l|l||ll|l|l|l|}
\caption[List of Wyckoff positions and BR dimensions in Type-IV SSGs with SOC]{List of Wyckoff positions (second column) in Type-IV SSGs with SOC (first column). Maximal Wyckoff positions are typed in bold. The third and fourth columns give the multiplicity of the Wyckoff position in a primitive unit cell and the dimensions of the BRs induced from the Wyckoff position with SOC. Wyckoff position that share the same values of these two parameters are grouped together. } \\
\hline
\multicolumn{2}{|c|}{Magnetic group}&Wyckoff pos.&m&d&\multicolumn{2}{|c|}{Magnetic group}&Wyckoff pos.&m&d\\
\hline
\endfirsthead
\hline
\multicolumn{10}{|c|}{List of Wyckoff positions and BR dimensions in Type-IV SSGs with SOC}\\
\hline
\multicolumn{2}{|c|}{Magnetic group}&Wyckoff pos.&m&d&\multicolumn{2}{|c|}{Magnetic group}&Wyckoff pos.&m&d\\
\hline
\endhead
\rule{0pt}{10pt}
1.3&$P_S1$&$\mathbf{a}$&2&2&2.7&$P_S\bar{1}$&$\mathbf{a},\mathbf{c},\mathbf{d},\mathbf{e}$&2&2\\
&&&&&&&$\mathbf{b},\mathbf{f},\mathbf{g},\mathbf{h}$&2&4\\
&&&&&&&$i$&4&4\\
\hline
\rule{0pt}{10pt}
3.4&$P_a2$&$\mathbf{a},\mathbf{b},\mathbf{c},\mathbf{d}$&2&2&3.5&$P_b2$&$\mathbf{a},\mathbf{b},\mathbf{c},\mathbf{d}$&2&2\\
&&$e$&4&4&&&$e$&4&4\\
\hline
\rule{0pt}{10pt}
3.6&$P_C2$&$\mathbf{a},\mathbf{b}$&2&2&4.10&$P_a2_1$&$\mathbf{a}$&4&4\\
&&$c$&4&4&&&&&\\
\hline
\rule{0pt}{10pt}
4.11&$P_b2_1$&$\mathbf{a},\mathbf{b},\mathbf{c},\mathbf{d}$&2&2&4.12&$P_C2_1$&$\mathbf{a},\mathbf{b}$&2&2\\
\end{longtable}

\subsection{Filling enforced conditions for single (without SOC) Type-I SSGs}\label{fecTypeINoSOC}
\begin{longtable}{|l|l|l|c|c|}
\caption[Filling enforced conditions of single (without SOC) Type-I SSG]{Filling enforced conditions of single (without SOC) Type-I SSGs. For each SSG, we tabulate the set of WPs that must be empty in the 2nd column. In the 3rd column, at least one of the WPs must be occupied. If the WPs in the 3rd column are grouped in different subsets, one of the WPs in each  parentheses must be occupied. The 4th column tabulates the necessary condition for a material to be a MTQC topologically trivial insulator. The 5th column gives the corresponding filling condition for a MTQC topologically trivial insulator to be obstructed.} \\
\hline
SG&Empty&At least one Wyckoff&condition&condition\\
&&position occupied&MTQC trivial&FEOAI\\
&&&insulator&\\
\hline
\endfirsthead
\hline
\multicolumn{5}{|c|}{Filling enforced conditions of single (without SOC) Type-I SSG}\\
\hline
SG&Empty&At least one Wyckoff&condition&condition\\
&&position occupied&MTQC trivial&FEOAI\\
&&&insulator&\\
\hline
\endhead
\multirow{1}{*}{  1.1   $P1$}&\multicolumn{4}{c|}{No Filling Enforced OAI}\\
\hline
\multirow{1}{*}{  2.4   $P\bar{1}$}&$a,b,c,d,e,f,g,h$&$i$&$N_e=2\mathbb{N}$&$N_e=4\mathbb{N}+2$\\\cline{2-5}
\hline\multirow{1}{*}{  3.1   $P2$}&$a,b,c,d$&$e$&$N_e=2\mathbb{N}$&$N_e=4\mathbb{N}+2$\\\cline{2-5}
\hline\multirow{1}{*}{  4.7   $P2_1$}&\multicolumn{4}{c|}{No Filling Enforced OAI}\\
\hline
\multirow{1}{*}{  5.13  $C2$}&$a,b$&$c$&$N_e=2\mathbb{N}$&$N_e=4\mathbb{N}+2$\\\cline{2-5}
\hline\multirow{1}{*}{  6.18  $Pm$}&$a,b$&$c$&$N_e=2\mathbb{N}$&$N_e=4\mathbb{N}+2$\\\cline{2-5}
\hline\multirow{1}{*}{  7.24  $Pc$}&\multicolumn{4}{c|}{No Filling Enforced OAI}\\
\hline
\multirow{1}{*}{  8.32  $Cm$}&$a$&$b$&$N_e=2\mathbb{N}$&$N_e=4\mathbb{N}+2$\\
\end{longtable}

\subsection{Filling enforced conditions for single (without SOC) Type-II SSGs}
The filling enforced conditions for Type-II SSGs, which are paramagnetic, coincide with those for the 230 DSGs as tabulated in Table~\ref{fec_230DSG} of Section~\ref{fec_DSG}.

\subsection{Filling enforced conditions for single (without SOC) Type-III SSGs}\label{fecTypeIIINoSOC}
\begin{longtable}{|l|l|l|c|c|}
\caption[Filling enforced conditions of single (without SOC) Type-III SSGs]{Filling enforced conditions of single (without SOC) Type-III SSGs. For each SSG, we tabulate the set of WPs that must be empty in the 2nd column. In the 3rd column, at least one of the WPs must be occupied. If the WPs in the 3rd column are grouped in different subsets, one of the WPs in each  parentheses must be occupied. The 4th column tabulates the necessary condition for a material to be a MTQC topologically trivial insulator. The 5th column gives the corresponding filling condition for a MTQC topologically trivial insulator to be obstructed.} \\
\hline
SG&Empty&At least one Wyckoff&condition&condition\\
&&position occupied&MTQC trivial&FEOAI\\
&&&insulator&\\
\hline
\endfirsthead
\hline
\multicolumn{5}{|c|}{Filling enforced conditions of single (without SOC) Type-III SSGs}\\
\hline
SG&Empty&At least one Wyckoff&condition&condition\\
&&position occupied&MTQC trivial&FEOAI\\
&&&insulator&\\
\hline
\endhead
\multirow{1}{*}{  2.6   $P\bar{1}'$}&$a,b,c,d,e,f,g,h$&$i$&$N_e=2\mathbb{N}$&$N_e=4\mathbb{N}+2$\\\cline{2-5}
\hline\multirow{1}{*}{  3.3   $P2'$}&$a,b,c,d$&$e$&$N_e=2\mathbb{N}$&$N_e=4\mathbb{N}+2$\\\cline{2-5}
\hline\multirow{1}{*}{  4.9   $P2_1'$}&\multicolumn{4}{c|}{No Filling Enforced OAI}\\
\hline
\multirow{1}{*}{  5.15  $C2'$}&$a,b$&$c$&$N_e=2\mathbb{N}$&$N_e=4\mathbb{N}+2$\\\cline{2-5}
\hline\multirow{1}{*}{  6.20  $Pm'$}&$a,b$&$c$&$N_e=2\mathbb{N}$&$N_e=4\mathbb{N}+2$\\\cline{2-5}
\hline\multirow{1}{*}{  7.26  $Pc'$}&\multicolumn{4}{c|}{No Filling Enforced OAI}\\
\hline
\multirow{1}{*}{  8.34  $Cm'$}&$a$&$b$&$N_e=2\mathbb{N}$&$N_e=4\mathbb{N}+2$\\\cline{2-5}
\hline\multirow{1}{*}{  9.39  $Cc'$}&\multicolumn{4}{c|}{No Filling Enforced OAI}\\
\end{longtable}

\subsection{Filling enforced conditions for single (without SOC) Type-IV SSGs}\label{fecTypeIVNoSOC}
\begin{longtable}{|l|l|l|c|c|}
\caption[Filling enforced conditions of single (without SOC) Type-IV SSGs]{Filling enforced conditions of single (without SOC) Type-IV SSGs. For each SSG, we tabulate the set of WPs that must be empty in the 2nd column. In the 3rd column, at least one of the WPs must be occupied. If the WPs in the 3rd column are grouped in different subsets, one of the WPs in each  parentheses must be occupied. The 4th column tabulates the necessary condition for a material to be a MTQC topologically trivial insulator. The 5th column gives the corresponding filling condition for a MTQC topologically trivial insulator to be obstructed.} \\
\hline
SG&Empty&At least one Wyckoff&condition&condition\\
&&position occupied&MTQC trivial&FEOAI\\
&&&insulator&\\
\hline
\endfirsthead
\hline
\multicolumn{5}{|c|}{Filling enforced conditions of single (without SOC) Type-IV SSGs}\\
\hline
SG&Empty&At least one Wyckoff&condition&condition\\
&&position occupied&MTQC trivial&FEOAI\\
&&&insulator&\\
\hline
\endhead
\multirow{1}{*}{  1.3   $P_S1$}&\multicolumn{4}{c|}{No Filling Enforced OAI}\\
\hline
\multirow{1}{*}{  2.7   $P_S\bar{1}$}&$a,b,c,d,e,f,g,h$&$i$&$N_e=4\mathbb{N}$&$N_e=8\mathbb{N}+4$\\\cline{2-5}
\hline\multirow{1}{*}{  3.4   $P_a2$}&$a,b,c,d$&$e$&$N_e=4\mathbb{N}$&$N_e=8\mathbb{N}+4$\\\cline{2-5}
\hline\multirow{1}{*}{  3.5   $P_b2$}&$a,b,c,d$&$e$&$N_e=4\mathbb{N}$&$N_e=8\mathbb{N}+4$\\\cline{2-5}
\hline\multirow{1}{*}{  3.6   $P_C2$}&$a,b$&$c$&$N_e=4\mathbb{N}$&$N_e=8\mathbb{N}+4$\\\cline{2-5}
\hline\multirow{1}{*}{  4.10  $P_a2_1$}&\multicolumn{4}{c|}{No Filling Enforced OAI}\\
\hline
\multirow{1}{*}{  4.11  $P_b2_1$}&$a,b,c,d$&$e$&$N_e=4\mathbb{N}$&$N_e=8\mathbb{N}+4$\\\cline{2-5}
\hline\multirow{1}{*}{  4.12  $P_C2_1$}&$a,b$&$c$&$N_e=4\mathbb{N}$&$N_e=8\mathbb{N}+4$\\
\end{longtable}

\subsection{Filling enforced conditions for double (with SOC) Type-I SSGs}\label{fecTypeISOC}

\begin{longtable}{|l|l|l|c|c|}
\caption[Filling enforced conditions of double (with SOC) Type-I SSGs]{Filling enforced conditions of double (with SOC) Type-I SSGs. For each SSG, we tabulate the set of WPs that must be empty in the 2nd column. In the 3rd column, at least one of the WPs must be occupied. If the WPs in the 3rd column are grouped in different subsets, one of the WPs in each  parentheses must be occupied. The 4th column tabulates the necessary condition for a material to be a MTQC topologically trivial insulator. The 5th column gives the corresponding filling condition for a MTQC topologically trivial insulator to be obstructed.} \\
\hline
SG&Empty&At least one Wyckoff&condition&condition\\
&&position occupied&MTQC trivial&FEOAI\\
&&&insulator&\\
\hline
\endfirsthead
\hline
\multicolumn{5}{|c|}{Filling enforced conditions of double (with SOC) Type-I SSGs}\\
\hline
SG&Empty&At least one Wyckoff&condition&condition\\
&&position occupied&MTQC trivial&FEOAI\\
&&&insulator&\\
\hline
\endhead
\multirow{1}{*}{  1.1   $P1$}&\multicolumn{4}{c|}{No Filling Enforced OAI}\\
\hline
\multirow{1}{*}{  2.4   $P\bar{1}$}&\multicolumn{4}{c|}{No Filling Enforced OAI}\\
\hline
\multirow{1}{*}{  3.1   $P2$}&\multicolumn{4}{c|}{No Filling Enforced OAI}\\
\hline
\multirow{1}{*}{  4.7   $P2_1$}&\multicolumn{4}{c|}{No Filling Enforced OAI}\\
\hline
\multirow{1}{*}{  5.13  $C2$}&\multicolumn{4}{c|}{No Filling Enforced OAI}\\
\hline
\multirow{1}{*}{  6.18  $Pm$}&\multicolumn{4}{c|}{No Filling Enforced OAI}\\
\hline
\multirow{1}{*}{  7.24  $Pc$}&\multicolumn{4}{c|}{No Filling Enforced OAI}\\
\hline
\multirow{1}{*}{  8.32  $Cm$}&\multicolumn{4}{c|}{No Filling Enforced OAI}\\
\end{longtable}

\subsection{Filling enforced conditions for double (with SOC) Type-II SSGs}
The filling enforced conditions for Type-II DSGs, which are paramagnetic, coincide with those for the 230 DSGs as tabulated in Table~\ref{fec_230DSG} of Section~\ref{fec_DSG}.

\subsection{Filling enforced conditions for double (with SOC) Type-III SSGs}\label{fecTypeIIISOC}

\begin{longtable}{|l|l|l|c|c|}
\caption[Filling enforced conditions of double (with SOC) Type-III SSGs]{Filling enforced conditions of double (with SOC) Type-III SSGs. For each SSG, we tabulate the set of WPs that must be empty in the 2nd column. In the 3rd column, at least one of the WPs must be occupied. If the WPs in the 3rd column are grouped in different subsets, one of the WPs in each  parentheses must be occupied. The 4th column tabulates the necessary condition for a material too be a MTQC topologically trivial insulator. The 5th column gives the corresponding filling condition for a MTQC topologically trivial insulator to be obstructed.} \\
\hline
SG&Empty&At least one Wyckoff&condition&condition\\
&&position occupied&MTQC trivial&FEOAI\\
&&&insulator&\\
\hline
\endfirsthead
\hline
\multicolumn{5}{|c|}{Filling enforced conditions of double (with SOC) Type-III SSGs}\\
\hline
SG&Empty&At least one Wyckoff&condition&condition\\
&&position occupied&MTQC trivial&FEOAI\\
&&&insulator&\\
\hline
\endhead
\multirow{1}{*}{  2.6   $P\bar{1}'$}&\multicolumn{4}{c|}{No Filling Enforced OAI}\\
\hline
\multirow{1}{*}{  3.3   $P2'$}&\multicolumn{4}{c|}{No Filling Enforced OAI}\\
\hline
\multirow{1}{*}{  4.9   $P2_1'$}&\multicolumn{4}{c|}{No Filling Enforced OAI}\\
\hline
\multirow{1}{*}{  5.15  $C2'$}&\multicolumn{4}{c|}{No Filling Enforced OAI}\\
\hline
\multirow{1}{*}{  6.20  $Pm'$}&\multicolumn{4}{c|}{No Filling Enforced OAI}\\
\hline
\multirow{1}{*}{  7.26  $Pc'$}&\multicolumn{4}{c|}{No Filling Enforced OAI}\\
\hline
\multirow{1}{*}{  8.34  $Cm'$}&\multicolumn{4}{c|}{No Filling Enforced OAI}\\
\hline
\multirow{1}{*}{  9.39  $Cc'$}&\multicolumn{4}{c|}{No Filling Enforced OAI}\\
\end{longtable}

\subsection{Filling enforced conditions for double (with SOC) Type-IV SSGs}\label{fecTypeIVSOC}
\begin{longtable}{|l|l|l|c|c|}
\caption[Filling enforced conditions of double (with SOC) Type-IV SSGs]{Filling enforced conditions of double (with SOC) Type-IV SSGs. For each SSG, we tabulate the set of WPs that must be empty in the 2nd column. In the 3rd column, at least one of the WPs must be occupied. If the WPs in the 3rd column are grouped in different subsets, one of the WPs in each  parentheses must be occupied. The 4th column tabulates the necessary condition for a material to be a MTQC topologically trivial insulator. The 5th column gives the corresponding filling condition for a MTQC topologically trivial insulator to be obstructed.} \\
\hline
SG&Empty&At least one Wyckoff&condition&condition\\
&&position occupied&MTQC trivial&FEOAI\\
&&&insulator&\\
\hline
\endfirsthead
\hline
\multicolumn{5}{|c|}{Filling enforced conditions of double (with SOC) Type-IV SSGs}\\
\hline
SG&Empty&At least one Wyckoff&condition&condition\\
&&position occupied&MTQC trivial&FEOAI\\
&&&insulator&\\
\hline
\endhead
\multirow{1}{*}{  1.3   $P_S1$}&\multicolumn{4}{c|}{No Filling Enforced OAI}\\
\hline
\multirow{1}{*}{  2.7   $P_S\bar{1}$}&$a,c,d,e$&$b,f,g,h$&$N_e=2\mathbb{N}$&$N_e=4\mathbb{N}+2$\\\cline{2-5}
\hline\multirow{1}{*}{  3.4   $P_a2$}&\multicolumn{4}{c|}{No Filling Enforced OAI}\\
\hline
\multirow{1}{*}{  3.5   $P_b2$}&\multicolumn{4}{c|}{No Filling Enforced OAI}\\
\hline
\multirow{1}{*}{  3.6   $P_C2$}&\multicolumn{4}{c|}{No Filling Enforced OAI}\\
\hline
\multirow{1}{*}{  4.10  $P_a2_1$}&\multicolumn{4}{c|}{No Filling Enforced OAI}\\
\hline
\multirow{1}{*}{  4.11  $P_b2_1$}&\multicolumn{4}{c|}{No Filling Enforced OAI}\\
\hline
\multirow{1}{*}{  4.12  $P_C2_1$}&\multicolumn{4}{c|}{No Filling Enforced OAI}\\
\end{longtable}

\end{document}